\documentclass[aip,jcp,preprint]{revtex4-1}

\usepackage[final]{graphicx}
\usepackage[dvipsnames]{color}
\usepackage{dcolumn}
\usepackage{hhline}
\usepackage{amsmath}
\usepackage{appendix}
\usepackage{moreverb}
\usepackage{booktabs}
\DeclareGraphicsExtensions{.pdf,.eps}

\begin{document}

\title{Many-body Green's function $GW$ and Bethe-Salpeter study of the optical excitations in a paradigmatic model dipeptide}

%
%
%
%

\author{C. Faber,$^{1,2}$ P. Boulanger,$^{1}$ I. Duchemin,$^{2}$ C. Attaccalite,$^{1}$ X. Blase$^{1}$}

\affiliation{ 
$^1$Institut N\'{e}el, CNRS and Universit\'{e} Joseph Fourier,
B.P. 166, 38042 Grenoble Cedex 09, France. \\
$^2$INAC, SP2M/L$\_$sim, CEA cedex 09, 38054 Grenoble, France. \\}

\date{\today}

\begin{abstract}
We study within the many-body Green's function $GW$ and Bethe-Salpeter formalisms the excitation energies of a paradigmatic model dipeptide, focusing on
the four lowest-lying local and charge-transfer excitations. Our $GW$ calculations are performed at the self-consistent level, updating first
the quasiparticle energies, and further the single-particle wavefunctions within the static Coulomb-hole plus screened-exchange approximation to the $GW$
self-energy operator. Important level crossings, as compared to the starting Kohn-Sham LDA spectrum, are identified. Our final Bethe-Salpeter singlet excitation
energies are found to agree, within 0.07 eV, with CASPT2 reference data, except for one charge-transfer state where the discrepancy can be as large as 0.5 eV. Our
results agree best with LC-BLYP and CAM-B3LYP calculations with enhanced long-range exchange, with a 0.1 eV mean absolute error. This has been
achieved employing a parameter-free formalism applicable to metallic or insulating extended or finite systems.
\end{abstract}


\maketitle


\section{Introduction}

Charge-transfer (CT) excitations, i.e. the creation of an excited electron and hole with weakly overlapping spatial
distributions, are a crucial feature in donor-acceptor systems, important e.g. for photovoltaic applications.  Indeed, in
organic photovoltaic cells, the separation of the strongly bound photogenerated electron-hole pairs is believed to take
place at the donor-acceptor interface through an intermediate
CT excited state. \cite{Sariciftci_C60_Polymer, Schmidt-Mende, Nphotonics_Rev_PolymerSC}
However, the exact mechanisms leading to charge separation remain rather controversial, \cite{Bakulin12,Caruso12,Yost13} urging for computational quantum mechanical studies
 which allow an accurate exploration of local and CT excitations at various energies.

Treating CT excitations theoretically using \textit{ab initio} methods remains difficult.
Wavefunction-based quantum chemistry methods such as multiconfigurational techniques, e.g. complete active space second order perturbation 
theory (CASPT2) or multi-reference configuration interaction (MRCI),\cite{CASPT2_MRCI} yield accurate results, but they are 
computationally too demanding to treat systems with more than a few tens of atoms. At the density functional theory (DFT) level, 
constrained DFT formalisms \cite{VanVoorhis06,Ghosh10} have proven to be extremely efficient in providing a good description of the lowest-lying CT excitation
in rather large systems, but generalizing such techniques to higher excited states remains a difficult issue.
Further, excited states wavefunctions, needed to calculate e.g. transfer rates, are not available.  
Using time-dependent density functional theory (TDDFT), \cite{Runge_TDDFT,Marques06,Casida09} one obtains the entire excitation 
spectrum of systems significantly larger than that amenable to e.g. CASPT2 or MRCI approaches.  However, it fails in most instances in 
reproducing CT excitations when standard (semi)local functionals are used. \cite{Dreuw_Failure_TDDFT_CT, Dreuw_Failure_TDDFT_Bacteriochlorin,Tozer_Failure_TDDFT_CT} 
Such difficulties paved the way for the success of range-separated hybrid functionals \cite{Savin96, Savin97, Savin04, Scuseria06} that are precise for both local and CT excitations, 
\cite{Iikura_LRC_TDDFT, Tawada_LRC_TDDFT, Yanai04, Baer_LRC_TDDFT, Kronik_Rev_12} even though the transferability 
from one system to another of the parameters controlling the  short- and long-range exchange contributions remains a difficult issue. 
\cite{Stein_TDDFT_parameter, Wong08, Lange_TDDFT_parameter}

Recently, an alternative approach derived from many-body perturbation
 theory (MBPT) within a Green's function formalism, the so-called $GW$ 
\cite{Hedin,StrinatiGW1,StrinatiGW2,Hybertsen_Louie, GodbySchlueter_GW, GW_review_1,GW_review_2}
and Bethe-Salpeter (BSE) \cite{Sham_BSE,Hanke79,Strinati82,Strinati,Rohlfing98,Benedict98,Albrecht98}
formalisms, initially developed and extensively tested for bulk semiconductors, has been 
applied successfully to the problem of CT excitations in gas phase organic systems. An accuracy of 0.1-0.15 eV as compared to 
experiment could be obtained for small gas phase donor-acceptor systems combining acenes and acene derivatives with the tetracyanoethylene
 (TCNE) acceptor.  \cite{Blase_Attaccalite_BSE_11, Baumeier12} Further, a similar agreement with coupled cluster (CC2) calculations \cite{CC2}
was obtained for intramolecular CT excitations in a coumarin family of interest for dye-sensitized 
solar cells.\cite{Faber_Coumarin} The accuracy of the $GW$/BSE approach was demonstrated to be equivalent to that 
of TDDFT calculations with range-separated hybrid functionals and optimized parameters, but with a 
parameter-free formalism providing equivalent accuracy for extended and finite size systems. However,
 the number of $GW$/BSE studies of CT excitations in gas phase organic systems remains very scarce and much work is still needed to benchmark the approach on a large variety of molecules.

In the present study, we explore within the $GW$/BSE formalism a small - even though delicate - system, namely a model dipeptide 
based on the N-methylacetamide $(C_3H_7NO)$ molecule (see Fig.~\ref{wfnkohnsham}a). This system denotes one of the first
cases where large errors have been observed at the TDDFT level, triggering its study by a large variety of approaches, including 
CASPT2,\cite{Serrano98}  TDDFT with various (semi)local, hybrid or range-separated functionals 
\cite{Yanai04, Peach08} and also a Bethe-Salpeter study based on an "empirical" $GW$ approach.\cite{Rocca10}
Difficulties were encountered to reproduce the CASPT2 results \cite{Serrano98} with unusual discrepancies 
between the mentioned state-of-the-art techniques. Moreover, a very large sensitivity of CT excitation energies on the
chosen functional parameters within e.g. the same CAM-B3LYP TDDFT framework was observed.\cite{Yanai04, Peach08}

In this work, we emphasize in particular the effect of self-consistency within the $GW$ formalism, updating both quasiparticle energies
and further single-particle wavefunctions within the so-called self-consistent Coulomb-hole plus screened-exchange 
(COHSEX) static approximation to $GW$.\cite{Farid} 
Important level reorderings are observed, as compared to Kohn-Sham  DFT calculations with semilocal functionals, which leads
to important changes in the absorption spectrum. The effect of updating the wavefunctions within self-consistent COHSEX 
is shown to be more marginal. Our (singlet) excitation energies show an excellent agreement with existing CASPT2 calculations for most local
and CT excitations, with a maximum error of 0.07 eV, except for a CT state shown to be blue shifted
by up to 0.5 eV as compared to CASPT2. Overall, our results agree best with CAM-B3LYP calculations with an "enhanced"
long-range exchange ($\alpha+\beta=0.8$) contribution and the original LC-BLYP formulation, showing a maximum mean absolute error of 0.1 eV for both local and CT excitations.

\section{Methodology and technical details}

Developed in the mid-60s - and later extended at the \textit{ab initio} level in the mid-80s - for the study of the electronic properties of extended semiconductors
and insulators, the $GW$ formalism \cite{Hedin,StrinatiGW1,StrinatiGW2,Hybertsen_Louie, GodbySchlueter_GW, GW_review_1,GW_review_2}
is now starting to be applied to organic molecules in the gas phase in order to assess its merits and limitations.
\cite{Tiago05, Dori06, Palummo09, Marom11, Foerster11, Samsonidze11, Sharifzadeh12, Umari12, Marom12, Koerzdoerfer12, Pham13} The $GW$ approach aims at providing accurate
 quasiparticle energy levels, including the ionization energy and electronic affinity.
As a brief overview, we introduce the non-local and energy-dependent self-energy operator 
${\Sigma}({\bf r},{\bf r}';E)$ that represents the effect of exchange and correlation in a generalized
eigenvalue equation:

\begin{equation}
 \left( { - {\nabla}^2 \over 2} + V^{ion}({\bf r}) + V^{H}({\bf r}) \right) \phi_n({\bf r}) + 
\int d{\bf r} \; \Sigma({\bf r},{\bf r}';E_n)  \phi_n({\bf r}') = E_n \phi_n({\bf r}),
\end{equation}

\noindent where $V^{ion}$ and $V^{H}$ stand for the ionic and Hartree potential, respectively. The self-energy operator ${\Sigma}$
includes all interactions beyond the Hartree contribution. In the so-called $GW$ approximation, it is simplified to: 

\begin{eqnarray*}
 {\Sigma}({\bf r},{\bf r}';E) &=& {i \over 2\pi} \int d{\omega} e^{i{\omega}{0^+}} G({\bf r},{\bf r}';E+\omega) W({\bf r},{\bf r}';\omega), \\
  G({\bf r},{\bf r}';E)       &=& \sum_n { \phi_n({\bf r}) \phi_n^*({\bf r}') \over E - \varepsilon_n + {0^+} \times sgn(\varepsilon_n - E_F) }, \\
  W({\bf r},{\bf r}';\omega)  &=& \int d{\bf r}'' \epsilon^{-1}({\bf r},{\bf r}'';\omega) V^C({\bf r}'',{\bf r}'), \\
      &=& V^C({\bf r},{\bf r}') +  \iint d{\bf r}''d{\bf r}''' V^C({\bf r},{\bf r}'') \chi_0({\bf r}'',{\bf r}''') W({\bf r}''',{\bf r}'),
\end{eqnarray*}

\noindent with $G$ is the time-ordered single-particle Green's function and $W$ is the dynamically screened 
Coulomb potential.  $\phi_n$ and $\varepsilon_n$ are input single-particle eigenstates/eigenenergies, respectively, typically taken from DFT Kohn-Sham
calculations (see below). $E_F$ is the Fermi level and $V^C$ the bare Coulomb potential. $\epsilon^{-1}$ denotes
the inverse dynamical dielectric matrix, calculated here below within the random phase approximation, and $\chi_0$ is the
 independent-electron susceptibility. The infinitesimally small positive value $(0^+)$ is included when carrying out a Fourier transformation from time
 to frequency space to ensure convergence. For the sake of comparison, the non-local, but
 energy-independent (instantaneous), exchange Fock operator reads :  

\begin{equation}
  {\Sigma}^{X}({\bf r},{\bf r}') = {i \over 2\pi} \int d{\omega} e^{i{\omega}{0^+}}
            G^{HF}({\bf r},{\bf r}';\omega) V^C({\bf r},{\bf r}'),
\end{equation}

\noindent where $G^{HF}$ is  built from Hartree-Fock single particle eigenstates. 

Our $GW$ calculations are performed with the {\sc{Fiesta}} package,\cite{Blase_FIESTA_code,Faber_DNA,Blase_Attaccalite_BSE_11} a recently
developed Gaussian-basis implementation of the $GW$ and Bethe-Salpeter formalisms. Dynamical screening and correlations are explicitly treated
 using contour deformation techniques without any plasmon-pole approximation (for more details on the contour deformation approach used, see \onlinecite{Blase_FIESTA_code, FaridGW}).
 All non-local operators such as
 the independent-electron susceptibility $\chi_0$,
the bare and screened Coulomb potentials $V^C$ and $W$ and the self-energy are expressed in terms of a large auxiliary 
atom-centered Gaussian basis combined with standard resolution-of-the-identity techniques.\cite{Vahtras1993514,1367-2630-14-5-053020,RI-MP2}  All unoccupied states,
 as appearing in the Green's function and independent-electron susceptibility, are included in the summation over empty states.
The used auxiliary basis is composed of six primitive Gaussian functions ($e^{- \alpha r^{2}}$) per \textit{l}-channel, up to \textit{l}=2 orbitals for first row 
elements, with an even tempered distribution \cite{Cherkes09} of the localization coefficients ($\alpha$) ranging from 
$\alpha_{min}$=0.10 $\text{Bohr}^{-2}$ to $\alpha_{max}$=3.2 $\text{Bohr}^{-2}$.  

Our starting input eigenstates/energies $(\phi_n,\varepsilon_n)$ are taken from DFT Kohn-Sham calculations using the { \sc{Siesta}} DFT code
 \cite{Siesta} within the local density approximation (LDA)\cite{LDA_1, LDA_3, LDA_2} combined with standard
norm-conserving pseudopotentials.\cite{Pseudos_Troullier_Martins} A large triple-zeta plus double polarization basis (TZDP) is
used to converge the correlation contribution to the self-energy (see note \onlinecite{TZ2P,Dunning89} for details). The influence of the starting 
eigenstates onto the final quasiparticle energies has recently become a subject of study,  
\cite{Blase_FIESTA_code,Bruneval13}
starting with the observation that the standard "single-shot" perturbative $G_0W_0$ calculations, i.e. $GW$ calculations based on  Kohn-Sham LDA
or PBE eigenstates/energies, tend to underestimate the gap of organic molecules. \cite{Blase_FIESTA_code,Marom12,Bruneval13}
To remedy this problem, our $GW$ calculations are  performed in a partially self-consistent way, where the 
obtained quasiparticle energies are reinjected into the calculation of the time-ordered Green's function and the screened 
Coulomb potential $W$, whereas the wavefunctions are left unchanged.  As emphasized in several recent works,
this approach yields quasiparticle energies, \cite{Blase_FIESTA_code,Faber_DNA,Neaton12,Marom12,Evers12}
electron-phonon coupling constants \cite{Faber_C60,Ciuchi12}
and optical absorption spectra \cite{Blase_Attaccalite_BSE_11,Baumeier12}
in much better agreement with experiment than non-selfconsistent single-shot $G_0W_0$ calculations based on a starting LDA Kohn-Sham spectrum. 

To study the influence of the input wavefunctions on the quasiparticle energies, we introduce a scheme which is
widely used in the $GW$ community for extended solids, namely a fully-self-consistent approach where both eigenstates
and eigenfunctions are updated using the so-called static Coulomb-hole plus screened-exchange (COHSEX) approximation
to the self-energy operator (see Ref.~\onlinecite{Bruneval06} for details).
The resulting modified eigenstates are then used to perform  partially self-consistent $GW$ calculations updating
the quasiparticle energy levels, but freezing the self-consistent COHSEX wavefunctions. 

Subsequently to the calculation of the quasi-particle spectrum using the $GW$ formalism, the (screened) Coulomb interaction 
between excited electrons and holes can be taken into account within the Bethe-Salpeter (BSE) formalism.
\cite{Sham_BSE,Hanke79,Strinati82,Strinati}  Here, the neutral excitation energies are the eigenvalues of the 
following electron-hole Hamiltonian equation:

\begin{equation}
\left( \begin{array}{cc} R & C \\ -C^{*} &  -R^*  \end{array} \right)  .
\left( \begin{array}{c} \left[  \phi_a({\bf r}_e) \phi_i({\bf r}_h)  \right]
        \\ \left[  \phi_i({\bf r}_e) \phi_a({\bf r}_h)  \right]  \end{array} \right)
 = \left( \begin{array}{c} \lambda_{ai} \\ \mu_{ia}  \end{array}  \right)
\left( \begin{array}{c} \left[ \phi_a({\bf r}_e) \phi_i({\bf r}_h)  \right]
        \\ \left[  \phi_i({\bf r}_e) \phi_a({\bf r}_h)  \right]   \end{array} \right),
\end{equation}

\noindent where the indexes (i,j) and (a,b) indicate the occupied and virtual orbitals, and
$({\bf r}_e,{\bf r}_h)$ the electron and hole positions, respectively.  In this block notation,
the vector $\left[ \phi_a({\bf r}_e) \phi_i({\bf r}_h) \right]$  represents all
excitations (note e.g. that $\phi_a({\bf r}_e)$ means that an electron  is put into a virtual orbital),
while the vector  $\left[ \phi_i({\bf r}_e) \phi_a({\bf r}_h) \right]$  represents all disexcitations.

The so-called resonant $R$ part is Hermitian 
and reads: 

\begin{multline}
   R_{ai,bj} = 
\delta_{a,b} \delta_{i,j} \left(  \varepsilon_a^{QP} - \varepsilon_i^{QP} \right) 
- \iint \phi_a( {\bf r}_e ) \phi_i({\bf r}_h) W({\bf r}_e,{\bf r}_h) \phi_{b}({\bf r}_e) \phi_{j}({\bf r}_h)  \, 
      \mathrm d{\bf r}_h \mathrm d{\bf r}_e \\
+ 2{\eta} \iint \phi_a( {\bf r}_e ) \phi_i({\bf r}_e) V^C({\bf r}_e,{\bf r}_h) \phi_{b}({\bf r}_h) \phi_{j}({\bf r}_h) \,
      \mathrm d{\bf r}_h \mathrm d{\bf r}_e,
\end{multline}

\noindent with $\eta=1$ for the singlet states studied here ($\eta=0$ for triplets). The quasiparticle energies $\varepsilon_{a,b,i,j}^{QP}$ are the $GW$ quasiparticle energies, 
while the $\phi_{a,b,i,j}$ are the Kohn-Sham eigenfunctions or the self-consistent COHSEX wavefunctions depending 
on the preceding $GW$ scheme (see above).  Notice that the electron-hole interaction term involving the screened Coulomb potential 
$W$ does not vanish for non-overlapping electron and hole states. In this limit, taking for sake of illustration
the case (i=j=h) and (a=b=l), where h and l stand for the HOMO and LUMO state, one obtains:

\begin{eqnarray*}
  R^{BSE}_{hl,hl} & \rightarrow & \left(  \varepsilon_{l}^{GW} -  \varepsilon_{h}^{GW} \right)
 -(1/\epsilon_M) \left< \;  |\phi_l({\bf r})|^2 |\phi_{h}({\bf r}')| ^2  V^C(|{\bf r} - {\bf r}'|) \; \right>,   \\
  R^{TDDFT}_{hl,hl} & \rightarrow & \left(  \varepsilon_{l}^{DFT} -  \varepsilon_{h}^{DFT} \right)
 -(\alpha+\beta) \left< \; |\phi_l({\bf r})|^2 |\phi_{h}({\bf r}')| ^2  V^C(|{\bf r} - {\bf r}'|) \;  \right>,
\end{eqnarray*}

\noindent where we included the TDDFT CAM-B3LYP expression \cite{Yanai04} for sake of comparison. These equations have been derived in the long-range limit,
 reducing $W(r,r')$ to $V^C(r,r')/\epsilon_M$, where $\epsilon_M$ is the macroscopic
dielectric constant, and setting to one the error function in the exact exchange contribution to the CAM-B3LYP functional. Clearly, through vacuum,
 where $\epsilon_M=$ 1, the correct asymptotic limit \cite{Mulliken}: ($IP-EA - 1/R$), with IP being the ionization potential, EA the electronic affinity and R an average
 HOMO to LUMO distance, is recovered. In the CAM-B3LYP case, the correct asymptotic limit would require an $\alpha+\beta=1$. We will come back to this point in the following.

Within the so-called Tamm-Dancoff approximation (TDA), the coupling between resonant ($R$) and anti-resonant ($R^*$) transitions
 is neglected, i.e. $C$ and $C^*$ are assumed to be zero. In the following, instead of applying the TDA, we diagonalize the full BSE matrix.
As recently shown in several BSE studies, \cite{Ma09,Gruening_TDA_09,Rocca10,Blase_Duchemin_BSE_12}
it is important to go beyond the TDA in nanosized systems, where it can lead 
to a blue-shift of the order of 0.3 eV. Finally, due to the quick increase in size of the  $(\phi_a \phi_i)$ product 
basis, we include all occupied states but restrict the contributing transitions to the lowest-lying 160 
unoccupied (virtual) states.~\cite{Note_NumberCondBands} 
The accuracy of the present $GW$/BSE formalism and its implementation has been tested recently in the case of small donor-acceptor 
complexes, with a mean absolute error (MAE) of 0.1-0.15 eV as compared to experiment for low-lying excitations showing 
a clear CT character.\cite{Blase_Attaccalite_BSE_11} It has been demonstrated that the results obtained with the present Gaussian basis 
implementation agree extremely well with planewave based $GW$/BSE calculations performed with 
the well established QuantumEspresso and Yambo code.\cite{QuantumEspresso,Yambo}  Similarly, intramolecular CT 
excitations in a family of coumarins displayed a MAE within 0.06 eV as compared to coupled cluster (CC2) calculations.\cite{Faber_Coumarin} 
We now address the delicate case of the dipeptide where, as shown 
below, significant differences have been observed between various methodologies.

\section{ Results and discussions}

\subsection{Notation}

\begin{figure}
\begin{center}
\includegraphics*[width=0.27\textwidth]{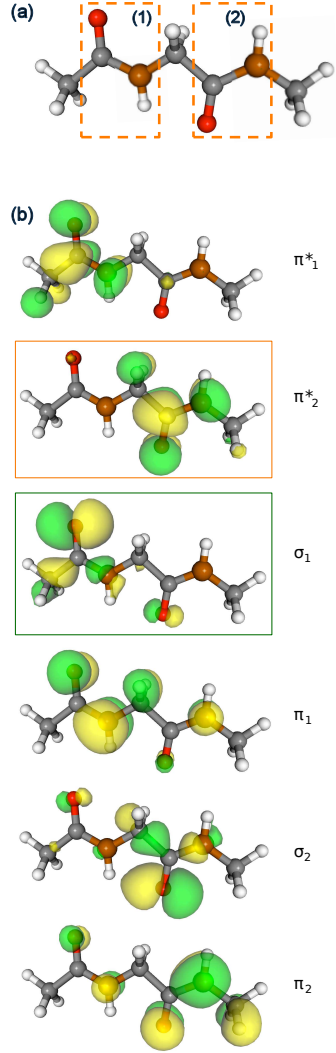}
\caption{ (Color online) a) Symbolic representation of the studied  model dipeptide, based on the N-methylacetamide $(C_3H_7NO)$ molecule, 
in its planar geometry (see Ref.~\onlinecite{Serrano98}, structure 1a). The active peptide bonds $(C(O)NH)$ are singled out in orange boxes.
 In CT excitations, the electron is promoted from one peptide group to the other. 
b) Isocontour representation of the Kohn-Sham wavefunctions around the gap classified by their $\sigma$- 
or $\pi$-character. The highest occupied molecular orbital (HOMO) and lowest unoccupied molecular orbital (LUMO) are singled 
out in a green and orange box, respectively. The asterisk denotes unoccupied orbitals, the subscripts (1, 2) the peptide unit on which the 
orbital is localized. The ordering corresponds to the DFT-LDA energy spectrum. 
Carbon atoms are represented in grey, oxygen in red, nitrogen in orange and hydrogen in white, respectively.} 
\label{wfnkohnsham}
\end{center}
\end{figure}

The model dipeptide studied here below was originally introduced in Ref.~\onlinecite{Serrano98} and studied subsequently
by a large variety of approaches \cite{Tozer99,Rocca10,Yanai04,Peach08,Akinaga09} as a test case for intramolecular
CT excitations. 
The molecular structure \cite{Structure_Fuelscher} is represented in Fig.~\ref{wfnkohnsham}a and was relaxed at a 
DFT-B3LYP \cite{B3lyp} level with a 6-311G(d,p) basis using the Gaussian09 package.\cite{Gaussian09, TZ2P}
The Kohn-Sham wavefunctions for states around the energy gap are depicted in Fig.~\ref{wfnkohnsham}b. It is primarily these 
states which contribute to the lowest-lying optical transitions. The listed states possess either $\sigma$-  or $\pi$- 
character and are mainly localized on one side of the molecule or the other (denoted with a subscript $i=1,2$). The asterisk 
stands for unoccupied (virtual) states. In this publication, we focus on two different kinds of excitations. On the one hand, we are interested
in valence transitions, labeled $W1$ and $W2$, between states 
localized on the same peptide unit, where the electron is promoted from an occupied $\sigma_i$-state to an unoccupied $\pi_i^*$-state.
On the other hand, we study CT excitations between states localized on different peptide groups, namely the $CTa$
exciton,  a $\sigma_1 \rightarrow \pi^*_2$ transition, and the $CTb$ exciton, a $\pi_1 \rightarrow \pi^*_2$ transition. 

\begin{figure}
\begin{center}
\includegraphics*[width=0.31\textwidth]{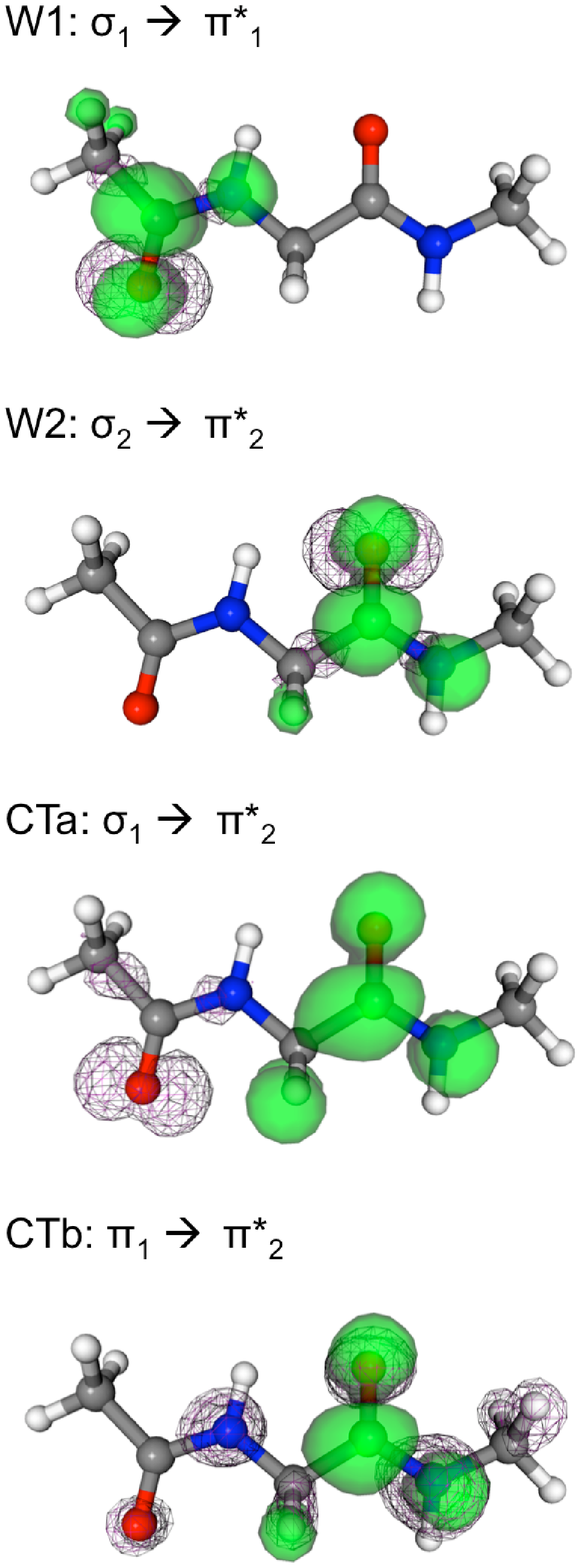}
\caption{ (Color online) Isocontour representation in green (wireframe) of the averaged electron (hole) distribution for 
the valence transitions $W_i$ (${\sigma}_i \rightarrow {\pi}_i$), the $CTa$ 
(${\sigma}_1 \rightarrow {\pi}_2^*$) and  the $CTb$ (${\pi}_1 \rightarrow {\pi}_2^*$) excitations. Carbon atoms are
represented in grey, oxygen in red, nitrogen in blue and hydrogen in white, respectively. }
\label{excitons}
\end{center}
\end{figure}

Anticipating on our $GW$/BSE results in section B and C, Fig.~\ref{excitons} illustrates the studied excitations by providing an isocontour representation of the hole-averaged 
electron distribution (transparent green) as obtained from the expectation value of the electron density operator 
$\delta({\bf r} - {\bf r}_e )$ on the corresponding two-body ${\psi}({\bf r}_e,{\bf r}_h)$ BSE eigenstate. Similarly, the electron-averaged hole distribution 
is represented (grey wireframe).  The very clear CT character of the $CTa$ transition and the partial 
CT character of the $CTb$ excitation can be easily verified.

\subsection{$GW$ and Bethe-Salpeter calculations beyond the scissor operator}

\begin{figure}
\begin{center}
\includegraphics*[width=0.4\textwidth]{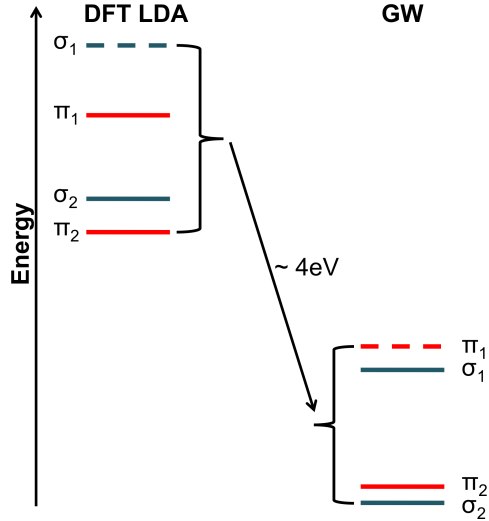}
\caption{ (Color online) Qualitative representation of the highest occupied molecular orbitals level ordering within DFT-LDA and $GW$. 
In blue the $\sigma$- states and in red the $\pi$- states. The HOMO is highlighted with a dashed line. Note the energy reordering 
and the changed spacing between the levels. }
\label{levelcrossing}
\end{center}
\end{figure}

As discussed in the technical details section, we first  perform self-consistency on the eigenvalues in our $GW$ 
calculations, while leaving the Kohn-Sham wavefunctions unchanged. As expected, the DFT-LDA 
Kohn-Sham energy  gap is significantly opened, from 4.6 eV to 11.9 eV within self-consistent $GW$. Beyond this known 
energy gap opening effect, important $\sigma$- and $\pi$-level crossings are observed for the highest occupied 
levels (see Fig.~\ref{levelcrossing}), with in particular a $GW$ correction about 0.5 eV larger for the two 
$\sigma$- states than for the two ${\pi}$- levels.  An important consequence is that the highest occupied molecular 
orbital (HOMO) changes its character from  $\sigma$ within DFT-LDA to $\pi$ within $GW$. One can speculate that the
stronger localization of the $\sigma$-orbitals leads to a larger 
self-interaction error as compared to $\pi$-orbitals, pushing them at too high energies at the DFT-LDA level.  A similar effect has been 
noticed in a recent $GW$ study on DNA/RNA nucleobases,\cite{Faber_DNA,Umari_DNA_11} where it has been shown that the present
partially self-consistent $GW$ scheme leads to quasiparticle energies in excellent agreement
with high-level quantum chemistry  \emph{ab initio} coupled-cluster and multiconfigurational perturbation methods such as CCSD(T), CASPT2 and EOM-IP-CCSD.

A recent study of the dipeptide, by Rocca and coworkers, \cite{Rocca10} used the BSE formalism directly on top of a DFT Kohn-Sham
calculation, for which the LDA HOMO-LUMO gap has been opened by hand using an "empirical" value.\cite{Note_scissor} This rigid shift preserved the DFT-LDA
ordering and energy spacing between occupied (unoccupied) orbitals, an approach labeled the "scissor" approximation to the $GW$ self-energy. 
We now analyze the results of our Bethe-Salpeter calculations starting from a quasiparticle ($GW$) spectrum presenting a corrected level 
spacing and ordering.

\subsection{Comparison to multi-reference quantum chemistry perturbation theory and TDDFT methods}

\begin{figure}
\begin{center}
\includegraphics*[width=0.45\textwidth]{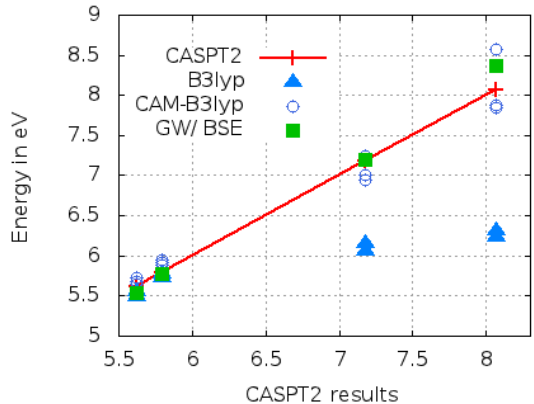}
\caption{ (Color online) Excitation energies as provided within several theoretical frameworks plotted against the CASPT2 results
(first diagonal in red). The TDDFT values with the B3LYP (blue up triangles) and the CAM-B3LYP (open circles) functionals, the 
present BSE calculations (green squares), starting from the $GW@$LDA partially self-consistent eigenstates (see text), are represented.
Several CAM-B3LYP values are found for each excitation, showing in particular the spread of values as a function of the ($\alpha+\beta$) 
parameter.  Energies are given in eV.}
\label{excitations}
\end{center}
\end{figure}

In Table~\ref{table1} and Fig.~\ref{excitations}, we provide the excitation energies as obtained by our $GW$/BSE calculations 
for the local $W1$ and $W2$ and the charge-transfer $CTa$ and $CTb$ transitions. Our $GW$/BSE values are compared to standard 
TDDFT-LDA calculations performed both with the {\sc{Fiesta}} code, using the same basis than for our BSE
calculations, and the QuantumEspresso package \cite{Rocca10,QuantumEspresso} using a planewave basis.  Further, the results of previous TDDFT calculations using the hybrid B3LYP, the long-range corrected
LC-BLYP and the Coulomb-attenuated CAM-B3LYP functionals with two different parametrisations \cite{Yanai04, Peach08} are 
presented, together with an early quantum chemistry CASPT2 calculation.~\cite{Serrano98}

Our TDDFT-LDA calculations come in very good agreement with the previous TDDFT-LDA planewave-based calculations
performed with the QuantumEspresso package.\cite{Rocca10} 
Both calculations predict CT states in nearly perfect agreement, with a negligible 0.02 eV discrepancy. The local $W1$ 
and $W2$ transitions agree within 0.1 eV. Such an agreement certainly comes as a good confirmation of the quality of the 
Kohn-Sham and auxiliary Gaussian bases used in the present study. Very similar results were also obtained at the TDDFT-PBE 
level in Ref.~\onlinecite{Peach08} with a maximum discrepancy of 0.05 eV as compared to our TDDFT-LDA calculations. 

The main outcome of the TDDFT-LDA or TDDFT-PBE calculations is that CT excitation energies are much too small. The
CT excitations are located below the lowest intramonomer $W1$ or $W2$ excitations. This is in great contrast to the CASPT2 
results, where the CT excitations are found to lie about 1.4 eV to 2.4 eV above the $W1$ and $W2$ transitions. Our 
TDDFT-LDA value  (4.63 eV) for the $CTa$ transition, which consists nearly entirely of a transition between the 
Kohn-Sham highest occupied (HOMO) and  lowest unoccupied (LUMO) molecular orbitals, can be compared to the HOMO-LUMO 
Kohn-Sham gap of 4.62 eV. This confirms that within TDDFT using local exchange-correlation functionals, the electron-hole 
interaction term vanishes for spatially separated electron and hole states and one is left with the energy difference 
between  Kohn-Sham states, neglecting any excitonic interaction.  On the other hand, the local $W1$ and $W2$ transitions, with 
a strong overlap between final and initial states, are much better described, even though showing a  0.2-0.3 eV red shift for the
$W1$ transition as compared to CASPT2.

Introducing some amount of exact exchange in addition to the charge-density-dependent TDDFT kernel yields a term similar to the BSE $W$ matrix 
elements, but with the  bare Coulomb potential $V^C$ instead of the screened Coulomb potential $W$. As a result, even 
non-overlapping electrons and holes can interact. Previous TDDFT-B3LYP calculations (see Table \ref{table1}) indeed show some improvement as compared 
to TDDFT-LDA by locating the CT states above the $W1$ and $W2$ transitions. However, compared to CASPT2 calculations,
the CT excitations energies are still about 1 eV to 1.8 eV too small, as a reminder that the B3LYP functional
captures only 20$\%$ of the exact Fock exchange operator. This problem can be cured using range-separated functionals 
such as LC-BLYP or CAM-B3LYP,
where the CT excitations come in much better agreement \cite{Yanai04,Peach08} with the quantum-chemistry reference as indicated in Table \ref{table1}. 
Nevertheless, within the CAM-B3LYP method itself, one observes energy differences in the order of 0.7 eV for the $CTa$
exciton, leading to the standard question of the proper choice of the needed parameters ($\alpha+\beta=0.65$ 
or $\alpha+\beta=0.8$ in the present case). This point will be discussed below.  

Comparing our $GW$/BSE calculations (@LDA column in Table \ref{table1}) to CASPT2 values, we find an excellent agreement for the $W1$, $W2$ and the 
$CTb$ exciton.  The maximum discrepancy is 0.07 eV for the $W1$ transition, while remarkably both the local $W2$ 
and charge-transfer $CTb$ excitation agree within 0.02 eV. \cite{Note_TDA2}  Clearly, tuning the $(\alpha,\beta)$ and range-seperation parameters may bring 
the CAM-B3LYP calculations in better agreement with CASPT2 values, but we emphasize that the present $GW$/BSE scheme
does not contain any adjustable parameters.  Concerning the oscillator strengths of the respective transitions,  the 
$GW$/BSE values are in reasonable agreement with the CASPT2 reference. The LC-BLYP and CAM-B3LYP values also agree 
for the transitions with vanishing oscillator strength, whereas they significantly underestimate the value of the oscillator strength for the $CTb$ 
exciton, where the $GW$/BSE oscillator strength is closer
to the CASPT2 value.  As observed recently in a $GW$/BSE study of intramolecular CT excitations in the 
coumarin family, \cite{Faber_Coumarin} obtaining an excellent agreement between the various formalisms proves more 
difficult for the oscillator strengths than for the corresponding excitation energies.

The largest discrepancy between the present $GW$/BSE@LDA and available CASPT2 calculations is of 0.3 eV for the 
$CTa$ excitation. For such a transition, our $GW$/BSE value is in nearly perfect agreement with the LC-BLYP
prediction,  lying in between the two CAM-B3LYP values.
As evidenced in Table I and Fig.~\ref{excitations}, observing  the rather large $\sim$0.7 eV variation between the two 
CAM-B3LYP values, such a transition is clearly very sensitive to the details of the exchange and correlation potential. 
Before commenting on such a deviation, we will test  the impact of using frozen Kohn-Sham LDA
eigenstates in the present $GW$ and Bethe-Salpeter approach here below.

\begin{center}
\begin{table*}
\small
\caption{ Singlet excitation energies for the model dipeptide as obtained within various TDDFT, many-body perturbation theory and CASPT2 approaches. 
Energies are in eV.  For the CAM-B3LYP columns, the (0.65) and (0.8) numbers indicate the $(\alpha+\beta)$ parameter that 
controls in particular the percentage of long-range exchange. The $@$LDA and $@$COHSEX columns indicate that the (partially)
self-consistent $GW$ calculations, with update of the self-consistent eigenvalues only, have been performed with either DFT-LDA
 or self-consistent COHSEX eigenstates as a starting point. Numbers in parenthesis are the oscillator strengths. Oscillator
strengths in the $(\alpha+\beta)=0.65$ CAM-B3LYP column are taken from Ref.~\onlinecite{Yanai04}. }
\begin{tabular}{l|cc|c|c|cc|cc|c}
\hline
           & \multicolumn{6}{c|}{TD-DFT}   & \multicolumn{2}{c|}{GW/BSE} & CASPT2$^d$  \\
\hline
           & \multicolumn{2}{c|}{LDA}  & B3LYP$^{b/c}$    & LC-BLYP & \multicolumn{2}{c|}{CAM-B3LYP}  & \multicolumn{2}{c|}{} \\
\hline

     & Ref$^{a}$ & Fiesta & Refs$^{b/c}$ & Ref.$^{b}$ & (0.65)$^{b/c}$ & (0.8)$^{b}$  &  $@$LDA    &  $@$COHSEX   &                \\ 

\hline
 $W1$   & 5.30  & 5.40  &  5.49/5.55 & 5.56 (.001) & 5.65/5.68 (.001) & 5.72 (.001) &    5.55 (.001) & 5.58 (.001) & 5.62 (.001)  \\
 $W2$   & 5.66  & 5.73  &  5.73/5.77 & 5.80 (.000) & 5.88/5.92 (.000) & 5.95 (.000) &    5.79 (.000) & 5.80 (.000) & 5.79 (.001)  \\
$CTb$   & 5.15  & 5.13  &  6.06/6.15 & 7.02 (.043) & 6.94/7.00 (.018) & 7.24 (.040) &    7.20 (.095) & 7.13 (.063) & 7.18 (.134)  \\
$CTa$   & 4.61  & 4.63  &  6.24/6.31 & 8.38 (.000) & 7.88/7.84 (.000) & 8.58 (.000) &    8.36 (.000) & 8.58 (.000) & 8.07 (.000)  \\
\hline
\end{tabular}

{\raggedright
$^{a}$Ref.~\onlinecite{Rocca10}   \\
$^{b}$Ref.~\onlinecite{Yanai04}   \\
$^{c}$Ref.~\onlinecite{Peach08}   \\
$^d$Ref.~\onlinecite{Serrano98} (Table 2, structure 1a). \\ }

\label{table1}
\end{table*}
\end{center}

\subsection{$GW$/BSE calculations starting from self-consistent COHSEX eigenstates}

In an attempt to better understand this delicate system and to explore the accuracy of the present $GW$/BSE formalism,
we finally test one of the common approximations in the $GW$ community, namely the assumption that the Kohn-Sham and quasiparticle
eigenfunctions strongly overlap, even though the energy gap may differ significantly. This has been demonstrated e.g. in the
case of bulk silicon in the early days of $GW$ calculations,\cite{Hybertsen_Louie} justifying the practice of updating 
the quasiparticle energies while freezing the starting Kohn-Sham orbitals.

A well-known example, where such an approximation fails, is the case of systems combining delocalized (\textit{s,p}) orbitals and tight 3\textit{d} 
levels such as transition metal oxides. Such a failure has been cured within a self-consistent $GW$ approach, where both 
quasiparticle energies and wavefunctions are updated.\cite{Faleev04,Shishkin07}
In the case of atoms or small molecular systems, it has been demonstrated recently that fully self-consistent $GW$ calculations 
\cite{Rostgaard10,Ke11,Bruneval12}
lead to better quasiparticle properties (ionization potential, HOMO-LUMO gap, etc.) than standard 
perturbative $G_0W_0$ calculations based on frozen Kohn-Sham LDA or PBE eigenstates.

Due to high computational costs for performing fully self-consistent $GW$ calculations, a scheme has been developed based on a simplified
"static" approximation to the $GW$ self-energy operator, the so-called static screened-exchange plus Coulomb-hole 
approximation already discussed in a early paper by Hedin,\cite{Hedin} where the $GW$ approach for the interacting 
homogeneous electron gas was introduced. In such an approach, full self-consistency with an update of both eigenvalues 
and eigenfunctions is performed at the COHSEX level, followed by a $GW$ calculation with self-consistency on the eigenvalues only.\cite{Bruneval06} Such a scheme, in the following labeled as $GW@$COHSEX, has been shown to yield excellent 
results in semiconductors combining extended and localized states,\cite{Bruneval06,Gatti07} as recently demonstrated in the case of 
transparent conductive oxides and quaternary thin films for photovoltaics \cite{Vidal10a,Vidal10b} or bulk gold.\cite{Rangel12} 
Very briefly, the two contributions to the COHSEX self-energy are:

\begin{eqnarray*}
  \Sigma^{SEX}({\bf r},{\bf r}') &=& - \sum_n^{occp} \phi_n({\bf r}) \phi_n^*({\bf r}') W({\bf r},{\bf r}'; \omega=0), \\
  \Sigma^{COH}({\bf r},{\bf r}') &=& {1 \over 2} W({\bf r},{\bf r}'; \omega=0) \delta({\bf r}-{\bf r}'),
\end{eqnarray*}

\noindent where the screened exchange $\Sigma^{SEX}$ term is analog to the bare exchange Fock operator - with a summation 
over states limited to the occupied manifold - but replacing the bare Coulomb potential by the screened one at zero frequency. 
The Coulomb hole $\Sigma^{COH}$ is a local operator with no summations over the eigenstates.
Such an approximation can be obtained by assuming that the poles of the inverse dielectric matrix are located at much 
higher energy than the typical electronic transition energies. Several demonstrations or interpretations have been proposed
in e.g. Refs.~\onlinecite{Hybertsen_Louie},~\onlinecite{Farid} and \onlinecite{Bruneval06}, including a time-domain analysis in the seminal paper by Lars Hedin.\cite{Hedin}

Our findings concerning the self-consistent COHSEX run are consistent with previous observations on extended 
semiconductors, \cite{Hybertsen_Louie} namely that the static COHSEX approximation overcorrects the energy gap. Our COHSEX 
HOMO-LUMO gap for the dipeptide is found to be 12.9 eV, instead of 4.62 eV within DFT-LDA and 11.8 eV for $GW@$COHSEX, 
respectively. Providing a first indication that updating the wavefunctions does not affect very significantly the quasiparticle energy spectrum,
we see that the 11.8 eV $GW@$COHSEX energy gap is in good agreement with the 11.9 eV $GW@$LDA value previously found. For the sake of comparison,
 the Hartree-Fock HOMO-LUMO gap is found to be 13.85 eV (all-electron cc-pVTZ Gaussian09 
value).  Clearly, the COHSEX gap is much closer to the final $GW$ value than the starting DFT-LDA HOMO-LUMO Kohn-Sham gap.  
As compared to $GW$ calculations, the slightly too large COHSEX gap originates mainly from the HOMO which is located nearly 
one eV too low in energy (overbinding), while the LUMO is found to agree within 0.1-0.2 eV with the final $GW$ value. 
\cite{LUMOvalue} 

Besides the improved value of the energy gap, an important finding is that the self-consistent 
COHSEX approximation yields the correct ordering of states. In particular, the HOMO level is the $\pi_1$ state, located 
(0.25,0.74,0.99) eV  above the $\sigma_1$, $\pi_2$ and $\sigma_2$ states, respectively, to be compared to spacings of (0.26,0.75,0.97) eV 
within the final $GW$@COHSEX value.  Such an excellent agreement in level ordering and energy spacing, together with a 
better HOMO-LUMO gap, indicates that the COHSEX energy spectrum is certainly a better starting point for $GW$ calculations as compared 
to the DFT-LDA Kohn-Sham Ansatz.

\begin{figure}
\begin{center}
\includegraphics*[width=0.45\textwidth]{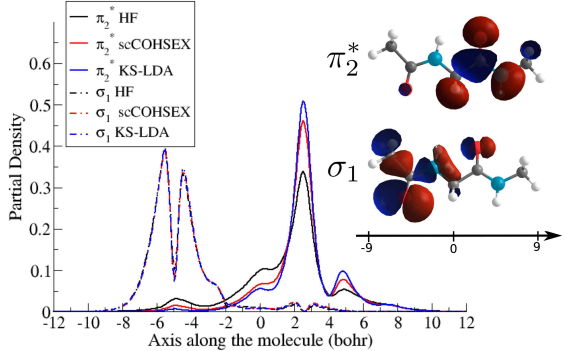}
\caption{ (Color online) Comparison between the LDA (blue), COHSEX (red) and HF (black) wavefunctions for the $\sigma_1$
(dotted lines) and $\pi_2^*$ (full lines)  wavefunctions.  We represent the modulus squared of the wavefunction averaged over 
planes perpendicular to the molecular ``axis" represented as an inset (average charge in electron/bohr). This partial density
averages to one for each state when integrated along the axis.  } 
\label{wfncohsex}
\end{center}
\end{figure}

Inferring a better quality of the COHSEX  eigenfunctions from the strongly ameliorated energy spectrum, as compared to
Kohn-Sham DFT-LDA calculations, remains a difficult issue. However, concerning the delicate $CTa$ transition, the analysis 
of the COHSEX $\sigma_1$ and $\pi_2^*$ states indicates that they project within 99.8$\%$ and 98.9$\%$, respectively, 
onto the corresponding LDA eigenstates. This shows that the Kohn-Sham and COHSEX eigenstates do not differ significantly, 
despite the very large difference in energy spectra.  For the sake of illustration,
we plot in Fig.~\ref{wfncohsex} the LDA, COHSEX and Hartree-Fock  $\sigma_1$ and $\pi_2^*$  wavefunctions 
averaging the charge within planes perpendicular to the molecular "axis". For the occupied $\sigma_1$ state,
the LDA,  COHSEX and Hartree-Fock wavefunctions (dotted lines) are nearly indistinguishable. However, for the
 $\pi_2^*$ state (full lines), differences start to appear in particular at the Hartree-Fock level. Clearly, the
COHSEX wavefunction is closer to the Kohn-Sham-LDA one, even though the COHSEX (and $GW$) quasiparticle spectrum 
is closer to the Hartree-Fock one.

The results of our $GW$/BSE study starting from self-consistent COHSEX eigenstates is presented in the column "@COHSEX" 
of Table~\ref{table1}. As compared to $GW$/BSE calculations where the Kohn-Sham eigenstates are kept frozen ("@LDA" column), 
the $W1$ and $W2$ excitation energies hardly change by a maximum of 0.03 eV for the $W1$ transition. The largest variation is again related to the $CTa$ transition,
with an increase of 0.22 eV, worsening the agreement with the CASPT2 value, but bringing our $GW$/BSE calculations in excellent 
agreement with the CAM-B3LYP ($\alpha+\beta=0.8$) results. Such an evolution can be traced back to a $\sim$ 0.2 eV blue-shift
of the $\pi_2^*$ energy level within $GW$@COHSEX as compared to $GW$@LDA. The oscillator strength associated with this transition 
is also seen to adopt a smaller value, worsening the agreement with the CASPT2 value, but improving the agreement with the 
CAM-B3LYP result.  

It is interesting to observe that what we may consider to be our most accurate values, namely our $GW$-Bethe-Salpeter
calculations based on the COHSEX eigenstates, come in excellent agreement with the CAM-B3LYP value with enhanced 
long-range exchange, namely setting ($\alpha+\beta$) to 0.8 instead of the original 0.65 value. We recall that in 
the case of CT excitations, the correct long-range "Mulliken" limit predicts a  $(-1/D)$  scaling
of the electron-hole binding energy, where $D$ is some measure of the donor to acceptor distance. Such a behavior
can only be reproduced with a ($\alpha+\beta=1$) parametrisation of the CAM-B3LYP functional. As such, the
($\alpha+\beta$=0.8) functional provides in principle a better description of the long-range CT 
electron-hole interaction. Very consistently,
the LC-BLYP functional, with a proper (-1/D) asymptotic scaling, locates the $CTa$ transition \cite{Yanai04} at 
8.38 eV, in much better agreement with our $GW$/BSE values than the CASPT2 prediction. However,
the analysis of the contributing wavefunctions in Fig.~\ref{wfncohsex} shows that the $CTa$ transition in the
dipeptide is far from the ideal case of the long-range well-separated electron-hole CT limit. Overall, our $GW$/BSE@COHSEX results
show a mean absolute error of 0.1 eV and 0.08 eV as compared to CAM-B3LYP ($\alpha+\beta=0.8$) and LC-BLYP, respectively. 

Regarding previous studies on CT excitations within the present $GW$/BSE formalism, with 
typical errors of the order of 0.1 eV as compared to experiment, TDDFT with optimized range-separated functionals 
or CASPT2 calculations, \cite{Blase_Attaccalite_BSE_11,Baumeier12,Faber_Coumarin} the present 0.3 eV to 0.5 eV 
discrepancies for the $CTa$ transition are somehow unusual, even though dramatically smaller than the typical 
errors induced by TDDFT calculation with semilocal kernels or even B3LYP. The 0.7 eV difference obtained 
between CAM-B3LYP calculations performed by the same authors with various parametrisations \cite{Yanai04} indicates that 
such variations cannot be explained by differences in running parameters (basis sizes and type, pseudopotential, etc.), 
but really hinges on the sensitivity of this transition onto the balance between short- and long-range exchange 
and correlation.

While we cannot comment on the accuracy and limitations of the available CASPT2 calculations, we certainly can  
emphasize in particular the lack of double-excitations in the present $GW$/BSE formalism and in TDDFT calculations, 
a possible explanation that would require more sophisticated treatments such as the inclusion of dynamical effects
in the screened Coulomb potential matrix elements at the BSE level. \cite{Sangalli11}
While this is certainly beyond the scope of the present paper, we can conclude that as it stands, the present 
parameter-free $GW$/BSE approach offers an accuracy comparable to TDDFT calculations performed 
with the best available parametrized range-separated functionals.

\section{Conclusions}

We studied within the many-body Green's function $GW$ and Bethe-Salpeter formalisms the excitation energies of
a paradigmatic dipeptide that has served  as a benchmark for describing intramolecular CT excitations
in organic systems within various theoretical frameworks, including TDDFT with local, global hybrid and range-separated hybrid functionals,
CASPT2 calculations and a previous Bethe-Salpeter study based on a model $GW$ approach. 
In the present work, we performed fully \textit{ab initio} $GW$ calculations, evidencing important $\sigma$/$\pi$ level reorderings
as compared to the starting Kohn-Sham LDA energy spectrum.  Based on $GW$ calculations with partial self-consistency on the quasiparticle
energies, our calculated Bethe-Salpeter excitation energies are found to agree with CASPT2 calculations with a discrepancy smaller than 0.07 eV for the local $W1$, $W2$
and charge-transfer $CTb$ excitations and a maximum discrepancy of 0.3 eV for the $CTa$ transition.
The effect of further updating self-consistently the quasiparticle wavefunctions within the static COHSEX approximation
to $GW$ leads to rather marginal variations for the $W1$, $W2$ and $CTb$ excitations, but shifts the discrepancy to
0.5 eV as compared to CASPT2 for the ubiquitous $CTa$ transition. In fine, our  BSE calculations based on the 
$GW@$COHSEX eigenvalues and eigenfunctions agree very well with both CAM-B3LYP calculations with enhanced 
long-range exchange ($\alpha+\beta=0.8$) and the original LC-BLYP formulation, with a maximum mean absolute error of 0.1 eV.
The present results allow to build  confidence in the use of the present parameter-free GW/BSE formalism in describing local
and charge-transfer excitations in organic systems of interest e.g. for photovoltaics, photosyntesis, or photocatalysis.


\section{Acknowledgments}  C.F. acknowledges a joint CEA/CNRS BDI fellowship and P.B. a postdoctoral fellowship from the French national research agency under
contract ANR-2012-BS04 PANELS. Computing time has been provided by the national 
GENGI-IDRIS supercomputing centers at Orsay under contract $n^o$ i2012096655 and a PRACE european project under contract 
$n^o$ 2012071258.  The authors are indebted to Prof. V. Robert and M. Verot for fruitful discussions concerning wavefunction 
based quantum chemistry approaches.


\begin{thebibliography}{100}
\makeatletter
\providecommand \@ifxundefined [1]{%
 \ifx #1\undefined \expandafter \@firstoftwo
 \else \expandafter \@secondoftwo
\fi
}%
\providecommand \@ifnum [1]{%
 \ifnum #1\expandafter \@firstoftwo
 \else \expandafter \@secondoftwo
\fi
}%
\providecommand \enquote [1]{``#1''}%
\providecommand \bibnamefont  [1]{#1}%
\providecommand \bibfnamefont [1]{#1}%
\providecommand \citenamefont [1]{#1}%
\providecommand\href[0]{\@sanitize\@href}%
\providecommand\@href[1]{\endgroup\@@startlink{#1}\endgroup\@@href}%
\providecommand\@@href[1]{#1\@@endlink}%
\providecommand \@sanitize [0]{\begingroup\catcode`\&12\catcode`\#12\relax}%
\@ifxundefined \pdfoutput {\@firstoftwo}{%
 \@ifnum{\z@=\pdfoutput}{\@firstoftwo}{\@secondoftwo}%
}{%
 \providecommand\@@startlink[1]{\leavevmode}%
 \providecommand\@@endlink[0]{}%
}{%
 \providecommand\@@startlink[1]{%
  \leavevmode
  \pdfstartlink
   attr{/Border[0 0 1 ]/H/I/C[0 1 1]}%
   user{/Subtype/Link/A<</Type/Action/S/URI/URI(#1)>>}%
  \relax
 }%
 \providecommand\@@endlink[0]{\pdfendlink}%
}%
\providecommand \url  [0]{\begingroup\@sanitize \@url }%
\providecommand \@url [1]{\endgroup\@href {#1}{\urlprefix}}%
\providecommand \urlprefix [0]{URL }%
\providecommand \Eprint[0]{\href }%
\@ifxundefined \urlstyle {%
  \providecommand \doi [1]{doi:\discretionary{}{}{}#1}%
}{%
  \providecommand \doi [0]{doi:\discretionary{}{}{}\begingroup
  \urlstyle{rm}\Url }%
}%
\providecommand \doibase [0]{http://dx.doi.org/}%
\providecommand \Doi[1]{\href{\doibase#1}}%
\providecommand \selectlanguage [0]{\@gobble}%
\providecommand \bibinfo [0]{\@secondoftwo}%
\providecommand \bibfield [0]{\@secondoftwo}%
\providecommand \translation [1]{[#1]}%
\providecommand \BibitemOpen[0]{}%
\providecommand \bibitemStop [0]{}%
\providecommand \bibitemNoStop [0]{.\EOS\space}%
\providecommand \EOS [0]{\spacefactor3000\relax}%
\providecommand \BibitemShut [1]{\csname bibitem#1\endcsname}%
\bibitem{Sariciftci_C60_Polymer}%
  \BibitemOpen
  \bibfield{author}{%
  \bibinfo {author} {\bibfnamefont{N.~S.}\ \bibnamefont{Sariciftci}}, \bibinfo
  {author} {\bibfnamefont{L.}~\bibnamefont{Smilowitz}}, \bibinfo {author}
  {\bibfnamefont{A.~J.}\ \bibnamefont{Heeger}},\ and\ \bibinfo {author}
  {\bibfnamefont{F.}~\bibnamefont{Wudl}},\ }%
  \bibfield{journal}{%
  \bibinfo {journal} {Science}\ }%
  \textbf{\bibinfo {volume} {258}},\ \bibinfo {pages} {1474} (\bibinfo {year}
  {1992})\BibitemShut{NoStop}%
\bibitem{Schmidt-Mende}%
  \BibitemOpen
  \bibfield{author}{%
  \bibinfo {author} {\bibfnamefont{L.}~\bibnamefont{Schmidt-Mende}}, \bibinfo
  {author} {\bibfnamefont{A.}~\bibnamefont{Fechtenkötter}}, \bibinfo {author}
  {\bibfnamefont{K.}~\bibnamefont{Müllen}}, \bibinfo {author}
  {\bibfnamefont{E.}~\bibnamefont{Moons}}, \bibinfo {author}
  {\bibfnamefont{R.~H.}\ \bibnamefont{Friend}},\ and\ \bibinfo {author}
  {\bibfnamefont{J.~D.}\ \bibnamefont{MacKenzie}},\ }%
  \bibfield{journal}{%
  \bibinfo {journal} {Science}\ }%
  \textbf{\bibinfo {volume} {293}},\ \bibinfo {pages} {1119} (\bibinfo {year}
  {2001})\BibitemShut{NoStop}%
\bibitem{Nphotonics_Rev_PolymerSC}%
  \BibitemOpen
  \bibfield{author}{%
  \bibinfo {author} {\bibfnamefont{G.}~\bibnamefont{Li}}, \bibinfo {author}
  {\bibfnamefont{R.}~\bibnamefont{Zhu}},\ and\ \bibinfo {author}
  {\bibfnamefont{Y.}~\bibnamefont{Yang}},\ }%
  \bibfield{journal}{%
  \bibinfo {journal} {Nat Photon}\ }%
  \textbf{\bibinfo {volume} {6}},\ \bibinfo {pages} {153} (\bibinfo {year}
  {2012})\BibitemShut{NoStop}%
\bibitem{Bakulin12}%
  \BibitemOpen
  \bibfield{author}{%
  \bibinfo {author} {\bibfnamefont{A.~A.}\ \bibnamefont{Bakulin}}, \bibinfo
  {author} {\bibfnamefont{A.}~\bibnamefont{Rao}}, \bibinfo {author}
  {\bibfnamefont{V.~G.}\ \bibnamefont{Pavelyev}}, \bibinfo {author}
  {\bibfnamefont{P.~H.~M.}\ \bibnamefont{van Loosdrecht}}, \bibinfo {author}
  {\bibfnamefont{M.~S.}\ \bibnamefont{Pshenichnikov}}, \bibinfo {author}
  {\bibfnamefont{D.}~\bibnamefont{Niedzialek}}, \bibinfo {author}
  {\bibfnamefont{J.}~\bibnamefont{Cornil}}, \bibinfo {author}
  {\bibfnamefont{D.}~\bibnamefont{Beljonne}},\ and\ \bibinfo {author}
  {\bibfnamefont{R.~H.}\ \bibnamefont{Friend}},\ }%
  \bibfield{journal}{%
  \bibinfo {journal} {Science}\ }%
  \textbf{\bibinfo {volume} {335}},\ \bibinfo {pages} {1340} (\bibinfo {year}
  {2012})\BibitemShut{NoStop}%
\bibitem{Caruso12}%
  \BibitemOpen
  \bibfield{author}{%
  \bibinfo {author} {\bibfnamefont{D.}~\bibnamefont{Caruso}}\ and\ \bibinfo
  {author} {\bibfnamefont{A.}~\bibnamefont{Troisi}},\ }%
  \bibfield{journal}{%
  \bibinfo {journal} {Proc. Natl. Acad. Sci.}\ }%
  \textbf{\bibinfo {volume} {109}},\ \bibinfo {pages} {13498} (\bibinfo {year}
  {2012})\BibitemShut{NoStop}%
\bibitem{Yost13}%
  \BibitemOpen
  \bibfield{author}{%
  \bibinfo {author} {\bibfnamefont{S.~R.}\ \bibnamefont{Yost}}\ and\ \bibinfo
  {author} {\bibfnamefont{T.}~\bibnamefont{Van~Voorhis}},\ }%
  \bibfield{journal}{%
  \bibinfo {journal} {J. Phys. Chem. C}\ }%
  \textbf{\bibinfo {volume} {117}},\ \bibinfo {pages} {5617} (\bibinfo {year}
  {2013})\BibitemShut{NoStop}%
\bibitem{CASPT2_MRCI}%
  \BibitemOpen
  \emph{\bibinfo {title} {Handbook of Computational Chemistry}},\ edited by\
  \bibinfo {editor} {\bibfnamefont{J.}~\bibnamefont{Leszczynski}}\ (\bibinfo
  {publisher} {Springer Verlag Berlin Heidelberg},\ \bibinfo {year}
  {2012})\BibitemShut{NoStop}%
\bibitem{VanVoorhis06}%
  \BibitemOpen
  \bibfield{author}{%
  \bibinfo {author} {\bibfnamefont{Q.}~\bibnamefont{Wu}}\ and\ \bibinfo
  {author} {\bibfnamefont{T.}~\bibnamefont{Van~Voorhis}},\ }%
  \bibfield{journal}{%
  \bibinfo {journal} {J. Chem. Theory Comput.}\ }%
  \textbf{\bibinfo {volume} {2}},\ \bibinfo {pages} {765} (\bibinfo {year}
  {2006})\BibitemShut{NoStop}%
\bibitem{Ghosh10}%
  \BibitemOpen
  \bibfield{author}{%
  \bibinfo {author} {\bibfnamefont{P.}~\bibnamefont{Ghosh}}\ and\ \bibinfo
  {author} {\bibfnamefont{R.}~\bibnamefont{Gebauer}},\ }%
  \bibfield{journal}{%
  \bibinfo {journal} {J. Chem. Phys.}\ }%
  \textbf{\bibinfo {volume} {132}},\ \bibinfo {pages} {104102} (\bibinfo {year}
  {2010})\BibitemShut{NoStop}%
\bibitem{Runge_TDDFT}%
  \BibitemOpen
  \bibfield{author}{%
  \bibinfo {author} {\bibfnamefont{E.}~\bibnamefont{Runge}}\ and\ \bibinfo
  {author} {\bibfnamefont{E.~K.~U.}\ \bibnamefont{Gross}},\ }%
  \bibfield{journal}{%
  \bibinfo {journal} {Phys. Rev. Lett.}\ }%
  \textbf{\bibinfo {volume} {52}},\ \bibinfo {pages} {997} (\bibinfo {year}
  {1984})\BibitemShut{NoStop}%
\bibitem{Marques06}%
  \BibitemOpen
  \bibfield{author}{%
  \bibinfo {author} {\bibfnamefont{M.~A.~L.}\ \bibnamefont{Marques}}, \bibinfo
  {author} {\bibfnamefont{C.}~\bibnamefont{Ullrich}}, \bibinfo {author}
  {\bibfnamefont{F.}~\bibnamefont{Nogueira}}, \bibinfo {author}
  {\bibfnamefont{A.}~\bibnamefont{Rubio}}, \bibinfo {author}
  {\bibfnamefont{K.}~\bibnamefont{Burke}},\ and\ \bibinfo {author}
  {\bibfnamefont{E.~K.~U.}\ \bibnamefont{Gross}},\ }%
  \emph{\bibinfo {title} {Time-Dependent Density Functional Theory}}\ (\bibinfo
  {publisher} {Springer Verlag Berlin Heidelberg},\ \bibinfo {year}
  {2006})\BibitemShut{NoStop}%
\bibitem{Casida09}%
  \BibitemOpen
  \bibfield{author}{%
  \bibinfo {author} {\bibfnamefont{M.~E.}\ \bibnamefont{Casida}},\ }%
  \bibfield{journal}{%
  \bibinfo {journal} {J. Mol. Struct.: THEOCHEM}\ }%
  \textbf{\bibinfo {volume} {914}},\ \bibinfo {pages} {3 } (\bibinfo {year}
  {2009})\BibitemShut{NoStop}%
\bibitem{Dreuw_Failure_TDDFT_CT}%
  \BibitemOpen
  \bibfield{author}{%
  \bibinfo {author} {\bibfnamefont{A.}~\bibnamefont{Dreuw}}, \bibinfo {author}
  {\bibfnamefont{J.~L.}\ \bibnamefont{Weisman}},\ and\ \bibinfo {author}
  {\bibfnamefont{M.}~\bibnamefont{Head-Gordon}},\ }%
  \bibfield{journal}{%
  \bibinfo {journal} {J. Chem. Phys.}\ }%
  \textbf{\bibinfo {volume} {119}},\ \bibinfo {pages} {2943} (\bibinfo {year}
  {2003})\BibitemShut{NoStop}%
\bibitem{Dreuw_Failure_TDDFT_Bacteriochlorin}%
  \BibitemOpen
  \bibfield{author}{%
  \bibinfo {author} {\bibfnamefont{A.}~\bibnamefont{Dreuw}}\ and\ \bibinfo
  {author} {\bibfnamefont{M.}~\bibnamefont{Head-Gordon}},\ }%
  \bibfield{journal}{%
  \bibinfo {journal} {J. Am. Chem. Soc.}\ }%
  \textbf{\bibinfo {volume} {126}},\ \bibinfo {pages} {4007} (\bibinfo {year}
  {2004})\BibitemShut{NoStop}%
\bibitem{Tozer_Failure_TDDFT_CT}%
  \BibitemOpen
  \bibfield{author}{%
  \bibinfo {author} {\bibfnamefont{D.~J.}\ \bibnamefont{Tozer}},\ }%
  \bibfield{journal}{%
  \bibinfo {journal} {J. Chem. Phys.}\ }%
  \textbf{\bibinfo {volume} {119}},\ \bibinfo {pages} {12697} (\bibinfo {year}
  {2003})\BibitemShut{NoStop}%
\bibitem{Savin96}%
  \BibitemOpen
  \bibfield{author}{%
  \bibinfo {author} {\bibfnamefont{A.}~\bibnamefont{Savin}},\ }%
  in\ \emph{\bibinfo {booktitle} {Recent Developments and Applications of
  Modern Density Functional Theory}},\ \bibinfo {editor} {edited by\ \bibinfo
  {editor} {\bibfnamefont{J.~M.}\ \bibnamefont{Seminario}}}\ (\bibinfo
  {publisher} {Elsevier},\ \bibinfo {year} {1996})\BibitemShut{NoStop}%
\bibitem{Savin97}%
  \BibitemOpen
  \bibfield{author}{%
  \bibinfo {author} {\bibfnamefont{T.}~\bibnamefont{Leininger}}, \bibinfo
  {author} {\bibfnamefont{H.}~\bibnamefont{Stoll}}, \bibinfo {author}
  {\bibfnamefont{H.-J.}\ \bibnamefont{Werner}},\ and\ \bibinfo {author}
  {\bibfnamefont{A.}~\bibnamefont{Savin}},\ }%
  \bibfield{journal}{%
  \bibinfo {journal} {Chem. Phys. Lett.}\ }%
  \textbf{\bibinfo {volume} {275}},\ \bibinfo {pages} {151 } (\bibinfo {year}
  {1997})\BibitemShut{NoStop}%
\bibitem{Savin04}%
  \BibitemOpen
  \bibfield{author}{%
  \bibinfo {author} {\bibfnamefont{J.}~\bibnamefont{Toulouse}}, \bibinfo
  {author} {\bibfnamefont{F.}~\bibnamefont{Colonna}},\ and\ \bibinfo {author}
  {\bibfnamefont{A.}~\bibnamefont{Savin}},\ }%
  \bibfield{journal}{%
  \bibinfo {journal} {Phys. Rev. A}\ }%
  \textbf{\bibinfo {volume} {70}},\ \bibinfo {pages} {062505} (\bibinfo {year}
  {2004})\BibitemShut{NoStop}%
\bibitem{Scuseria06}%
  \BibitemOpen
  \bibfield{author}{%
  \bibinfo {author} {\bibfnamefont{O.~A.}\ \bibnamefont{Vydrov}}, \bibinfo
  {author} {\bibfnamefont{J.}~\bibnamefont{Heyd}}, \bibinfo {author}
  {\bibfnamefont{A.~V.}\ \bibnamefont{Krukau}},\ and\ \bibinfo {author}
  {\bibfnamefont{G.~E.}\ \bibnamefont{Scuseria}},\ }%
  \bibfield{journal}{%
  \bibinfo {journal} {The Journal of Chemical Physics}\ }%
  \textbf{\bibinfo {volume} {125}},\ \bibinfo {pages} {074106} (\bibinfo {year}
  {2006})\BibitemShut{NoStop}%
\bibitem{Iikura_LRC_TDDFT}%
  \BibitemOpen
  \bibfield{author}{%
  \bibinfo {author} {\bibfnamefont{H.}~\bibnamefont{Iikura}}, \bibinfo {author}
  {\bibfnamefont{T.}~\bibnamefont{Tsuneda}}, \bibinfo {author}
  {\bibfnamefont{T.}~\bibnamefont{Yanai}},\ and\ \bibinfo {author}
  {\bibfnamefont{K.}~\bibnamefont{Hirao}},\ }%
  \bibfield{journal}{%
  \bibinfo {journal} {J. Chem. Phys.}\ }%
  \textbf{\bibinfo {volume} {115}},\ \bibinfo {pages} {3540} (\bibinfo {year}
  {2001})\BibitemShut{NoStop}%
\bibitem{Tawada_LRC_TDDFT}%
  \BibitemOpen
  \bibfield{author}{%
  \bibinfo {author} {\bibfnamefont{Y.}~\bibnamefont{Tawada}}, \bibinfo {author}
  {\bibfnamefont{T.}~\bibnamefont{Tsuneda}}, \bibinfo {author}
  {\bibfnamefont{S.}~\bibnamefont{Yanagisawa}}, \bibinfo {author}
  {\bibfnamefont{T.}~\bibnamefont{Yanai}},\ and\ \bibinfo {author}
  {\bibfnamefont{K.}~\bibnamefont{Hirao}},\ }%
  \bibfield{journal}{%
  \bibinfo {journal} {J. Chem. Phys.}\ }%
  \textbf{\bibinfo {volume} {120}},\ \bibinfo {pages} {8425} (\bibinfo {year}
  {2004})\BibitemShut{NoStop}%
\bibitem{Yanai04}%
  \BibitemOpen
  \bibfield{author}{%
  \bibinfo {author} {\bibfnamefont{T.}~\bibnamefont{Yanai}}, \bibinfo {author}
  {\bibfnamefont{D.~P.}\ \bibnamefont{Tew}},\ and\ \bibinfo {author}
  {\bibfnamefont{N.~C.}\ \bibnamefont{Handy}},\ }%
  \bibfield{journal}{%
  \bibinfo {journal} {Chem. Phys. Lett.}\ }%
  \textbf{\bibinfo {volume} {393}},\ \bibinfo {pages} {51 } (\bibinfo {year}
  {2004})\BibitemShut{NoStop}%
\bibitem{Baer_LRC_TDDFT}%
  \BibitemOpen
  \bibfield{author}{%
  \bibinfo {author} {\bibfnamefont{R.}~\bibnamefont{Baer}}\ and\ \bibinfo
  {author} {\bibfnamefont{D.}~\bibnamefont{Neuhauser}},\ }%
  \bibfield{journal}{%
  \bibinfo {journal} {Phys. Rev. Lett.}\ }%
  \textbf{\bibinfo {volume} {94}},\ \bibinfo {pages} {043002} (\bibinfo {month}
  {Feb}\ \bibinfo {year} {2005})\BibitemShut{NoStop}%
\bibitem{Kronik_Rev_12}%
  \BibitemOpen
  \bibfield{author}{%
  \bibinfo {author} {\bibfnamefont{L.}~\bibnamefont{Kronik}}, \bibinfo {author}
  {\bibfnamefont{T.}~\bibnamefont{Stein}}, \bibinfo {author}
  {\bibfnamefont{S.}~\bibnamefont{Refaely-Abramson}},\ and\ \bibinfo {author}
  {\bibfnamefont{R.}~\bibnamefont{Baer}},\ }%
  \bibfield{journal}{%
  \bibinfo {journal} {J. Chem. Theory Comput.}\ }%
  \textbf{\bibinfo {volume} {8}},\ \bibinfo {pages} {1515} (\bibinfo {year}
  {2012})\BibitemShut{NoStop}%
\bibitem{Stein_TDDFT_parameter}%
  \BibitemOpen
  \bibfield{author}{%
  \bibinfo {author} {\bibfnamefont{T.}~\bibnamefont{Stein}}, \bibinfo {author}
  {\bibfnamefont{L.}~\bibnamefont{Kronik}},\ and\ \bibinfo {author}
  {\bibfnamefont{R.}~\bibnamefont{Baer}},\ }%
  \bibfield{journal}{%
  \bibinfo {journal} {J. Am. Chem. Soc.}\ }%
  \textbf{\bibinfo {volume} {131}},\ \bibinfo {pages} {2818} (\bibinfo {year}
  {2009})\BibitemShut{NoStop}%
\bibitem{Wong08}%
  \BibitemOpen
  \bibfield{author}{%
  \bibinfo {author} {\bibfnamefont{B.~M.}\ \bibnamefont{Wong}}\ and\ \bibinfo
  {author} {\bibfnamefont{J.~G.}\ \bibnamefont{Cordaro}},\ }%
  \bibfield{journal}{%
  \bibinfo {journal} {J. Chem. Phys.}\ }%
  \textbf{\bibinfo {volume} {129}},\ \bibinfo {pages} {214703} (\bibinfo {year}
  {2008})\BibitemShut{NoStop}%
\bibitem{Lange_TDDFT_parameter}%
  \BibitemOpen
  \bibfield{author}{%
  \bibinfo {author} {\bibfnamefont{A.~W.}\ \bibnamefont{Lange}}, \bibinfo
  {author} {\bibfnamefont{M.~A.}\ \bibnamefont{Rohrdanz}},\ and\ \bibinfo
  {author} {\bibfnamefont{J.~M.}\ \bibnamefont{Herbert}},\ }%
  \bibfield{journal}{%
  \bibinfo {journal} {J. Phys. Chem. B}\ }%
  \textbf{\bibinfo {volume} {112}},\ \bibinfo {pages} {6304} (\bibinfo {year}
  {2008})\BibitemShut{NoStop}%
\bibitem{Hedin}%
  \BibitemOpen
  \bibfield{author}{%
  \bibinfo {author} {\bibfnamefont{L.}~\bibnamefont{Hedin}},\ }%
  \bibfield{journal}{%
  \bibinfo {journal} {Phys. Rev.}\ }%
  \textbf{\bibinfo {volume} {139}},\ \bibinfo {pages} {A796} (\bibinfo {year}
  {1965})\BibitemShut{NoStop}%
\bibitem{StrinatiGW1}%
  \BibitemOpen
  \bibfield{author}{%
  \bibinfo {author} {\bibfnamefont{G.}~\bibnamefont{Strinati}}, \bibinfo
  {author} {\bibfnamefont{H.}~\bibnamefont{Mattausch}},\ and\ \bibinfo {author}
  {\bibfnamefont{W.}~\bibnamefont{Hanke}},\ }%
  \bibfield{journal}{%
  \bibinfo {journal} {Phys. Rev. Lett.}\ }%
  \textbf{\bibinfo {volume} {45}},\ \bibinfo {pages} {290} (\bibinfo {year}
  {1980})\BibitemShut{NoStop}%
\bibitem{StrinatiGW2}%
  \BibitemOpen
  \bibfield{author}{%
  \bibinfo {author} {\bibfnamefont{G.}~\bibnamefont{Strinati}}, \bibinfo
  {author} {\bibfnamefont{H.}~\bibnamefont{Mattausch}},\ and\ \bibinfo {author}
  {\bibfnamefont{W.}~\bibnamefont{Hanke}},\ }%
  \bibfield{journal}{%
  \bibinfo {journal} {Phys. Rev. B}\ }%
  \textbf{\bibinfo {volume} {25}},\ \bibinfo {pages} {2867} (\bibinfo {year}
  {1982})\BibitemShut{NoStop}%
\bibitem{Hybertsen_Louie}%
  \BibitemOpen
  \bibfield{author}{%
  \bibinfo {author} {\bibfnamefont{M.~S.}\ \bibnamefont{Hybertsen}}\ and\
  \bibinfo {author} {\bibfnamefont{S.~G.}\ \bibnamefont{Louie}},\ }%
  \bibfield{journal}{%
  \bibinfo {journal} {Phys. Rev. B}\ }%
  \textbf{\bibinfo {volume} {34}},\ \bibinfo {pages} {5390} (\bibinfo {year}
  {1986})\BibitemShut{NoStop}%
\bibitem{GodbySchlueter_GW}%
  \BibitemOpen
  \bibfield{author}{%
  \bibinfo {author} {\bibfnamefont{R.~W.}\ \bibnamefont{Godby}}, \bibinfo
  {author} {\bibfnamefont{M.}~\bibnamefont{Schl\"uter}},\ and\ \bibinfo
  {author} {\bibfnamefont{L.~J.}\ \bibnamefont{Sham}},\ }%
  \bibfield{journal}{%
  \bibinfo {journal} {Phys. Rev. B}\ }%
  \textbf{\bibinfo {volume} {37}},\ \bibinfo {pages} {10159} (\bibinfo {month}
  {Jun}\ \bibinfo {year} {1988})\BibitemShut{NoStop}%
\bibitem{GW_review_1}%
  \BibitemOpen
  \bibfield{author}{%
  \bibinfo {author} {\bibfnamefont{F.}~\bibnamefont{Aryasetiawan}}\ and\
  \bibinfo {author} {\bibfnamefont{O.}~\bibnamefont{Gunnarsson}},\ }%
  \bibfield{journal}{%
  \bibinfo {journal} {Rep. Prog. Phys.}\ }%
  \textbf{\bibinfo {volume} {61}},\ \bibinfo {pages} {237} (\bibinfo {year}
  {1998})\BibitemShut{NoStop}%
\bibitem{GW_review_2}%
  \BibitemOpen
  \bibfield{author}{%
  \bibinfo {author} {\bibfnamefont{W.~G.}\ \bibnamefont{Aulbur}}, \bibinfo
  {author} {\bibfnamefont{L.}~\bibnamefont{J\"onsson}},\ and\ \bibinfo {author}
  {\bibfnamefont{J.~W.}\ \bibnamefont{Wilkins}},\ }%
  in\ \emph{\bibinfo {booktitle} {Solid State Physics}},\ \bibinfo {editor}
  {edited by\ \bibinfo {editor} {\bibfnamefont{H.}~\bibnamefont{Ehrenreich}}\
  and\ \bibinfo {editor} {\bibfnamefont{F.}~\bibnamefont{Spaepen}}}\ (\bibinfo
  {publisher} {Academic Press},\ \bibinfo {year} {1999})\BibitemShut{NoStop}%
\bibitem{Sham_BSE}%
  \BibitemOpen
  \bibfield{author}{%
  \bibinfo {author} {\bibfnamefont{L.~J.}\ \bibnamefont{Sham}}\ and\ \bibinfo
  {author} {\bibfnamefont{T.~M.}\ \bibnamefont{Rice}},\ }%
  \bibfield{journal}{%
  \bibinfo {journal} {Phys. Rev.}\ }%
  \textbf{\bibinfo {volume} {144}},\ \bibinfo {pages} {708} (\bibinfo {month}
  {Apr}\ \bibinfo {year} {1966})\BibitemShut{NoStop}%
\bibitem{Hanke79}%
  \BibitemOpen
  \bibfield{author}{%
  \bibinfo {author} {\bibfnamefont{W.}~\bibnamefont{Hanke}}\ and\ \bibinfo
  {author} {\bibfnamefont{L.~J.}\ \bibnamefont{Sham}},\ }%
  \bibfield{journal}{%
  \bibinfo {journal} {Phys. Rev. Lett.}\ }%
  \textbf{\bibinfo {volume} {43}},\ \bibinfo {pages} {387} (\bibinfo {year}
  {1979})\BibitemShut{NoStop}%
\bibitem{Strinati82}%
  \BibitemOpen
  \bibfield{author}{%
  \bibinfo {author} {\bibfnamefont{G.}~\bibnamefont{Strinati}},\ }%
  \bibfield{journal}{%
  \bibinfo {journal} {Phys. Rev. Lett.}\ }%
  \textbf{\bibinfo {volume} {49}},\ \bibinfo {pages} {1519} (\bibinfo {year}
  {1982})\BibitemShut{NoStop}%
\bibitem{Strinati}%
  \BibitemOpen
  \bibfield{author}{%
  \bibinfo {author} {\bibfnamefont{G.}~\bibnamefont{Strinati}},\ }%
  \bibfield{journal}{%
  \bibinfo {journal} {Rivista del nuovo cimento}\ }%
  \textbf{\bibinfo {volume} {11}},\ \bibinfo {pages} {1} (\bibinfo {year}
  {1988})\BibitemShut{NoStop}%
\bibitem{Rohlfing98}%
  \BibitemOpen
  \bibfield{author}{%
  \bibinfo {author} {\bibfnamefont{M.}~\bibnamefont{Rohlfing}}\ and\ \bibinfo
  {author} {\bibfnamefont{S.~G.}\ \bibnamefont{Louie}},\ }%
  \bibfield{journal}{%
  \bibinfo {journal} {Phys. Rev. Lett.}\ }%
  \textbf{\bibinfo {volume} {80}},\ \bibinfo {pages} {3320} (\bibinfo {year}
  {1998})\BibitemShut{NoStop}%
\bibitem{Benedict98}%
  \BibitemOpen
  \bibfield{author}{%
  \bibinfo {author} {\bibfnamefont{L.~X.}\ \bibnamefont{Benedict}}, \bibinfo
  {author} {\bibfnamefont{E.~L.}\ \bibnamefont{Shirley}},\ and\ \bibinfo
  {author} {\bibfnamefont{R.~B.}\ \bibnamefont{Bohn}},\ }%
  \bibfield{journal}{%
  \bibinfo {journal} {Phys. Rev. Lett.}\ }%
  \textbf{\bibinfo {volume} {80}},\ \bibinfo {pages} {4514} (\bibinfo {year}
  {1998})\BibitemShut{NoStop}%
\bibitem{Albrecht98}%
  \BibitemOpen
  \bibfield{author}{%
  \bibinfo {author} {\bibfnamefont{S.}~\bibnamefont{Albrecht}}, \bibinfo
  {author} {\bibfnamefont{L.}~\bibnamefont{Reining}}, \bibinfo {author}
  {\bibfnamefont{R.}~\bibnamefont{Del~Sole}},\ and\ \bibinfo {author}
  {\bibfnamefont{G.}~\bibnamefont{Onida}},\ }%
  \bibfield{journal}{%
  \bibinfo {journal} {Phys. Rev. Lett.}\ }%
  \textbf{\bibinfo {volume} {80}},\ \bibinfo {pages} {4510} (\bibinfo {year}
  {1998})\BibitemShut{NoStop}%
\bibitem{Blase_Attaccalite_BSE_11}%
  \BibitemOpen
  \bibfield{author}{%
  \bibinfo {author} {\bibfnamefont{X.}~\bibnamefont{Blase}}\ and\ \bibinfo
  {author} {\bibfnamefont{C.}~\bibnamefont{Attaccalite}},\ }%
  \bibfield{journal}{%
  \bibinfo {journal} {Appl. Phys. Lett.}\ }%
  \textbf{\bibinfo {volume} {99}},\ \bibinfo {pages} {171909} (\bibinfo {month}
  {oct}\ \bibinfo {year} {2011})\BibitemShut{NoStop}%
\bibitem{Baumeier12}%
  \BibitemOpen
  \bibfield{author}{%
  \bibinfo {author} {\bibfnamefont{B.}~\bibnamefont{Baumeier}}, \bibinfo
  {author} {\bibfnamefont{D.}~\bibnamefont{Andrienko}}, \bibinfo {author}
  {\bibfnamefont{Y.}~\bibnamefont{Ma}},\ and\ \bibinfo {author}
  {\bibfnamefont{M.}~\bibnamefont{Rohlfing}},\ }%
  \bibfield{journal}{%
  \bibinfo {journal} {J. Chem. Theory Comput.}\ }%
  \textbf{\bibinfo {volume} {8}},\ \bibinfo {pages} {997} (\bibinfo {year}
  {2012})\BibitemShut{NoStop}%
\bibitem{CC2}%
  \BibitemOpen
  \emph{\bibinfo {title} {Recent Progress in Coupled Cluster Methods}},\ edited
  by\ \bibinfo {editor} {\bibfnamefont{P.}~\bibnamefont{C\'arsky}}, \bibinfo
  {editor} {\bibfnamefont{J.}~\bibnamefont{Paldus}},\ and\ \bibinfo {editor}
  {\bibfnamefont{J.}~\bibnamefont{Pittner}},\ \bibinfo {series} {Challenges and
  Advances in Computational Chemistry and Physics}, Vol.~\bibinfo {volume}
  {11}\ (\bibinfo {publisher} {Springer Verlag Berlin Heidelberg},\ \bibinfo
  {year} {2010})\BibitemShut{NoStop}%
\bibitem{Faber_Coumarin}%
  \BibitemOpen
  \bibfield{author}{%
  \bibinfo {author} {\bibfnamefont{C.}~\bibnamefont{Faber}}, \bibinfo {author}
  {\bibfnamefont{I.}~\bibnamefont{Duchemin}}, \bibinfo {author}
  {\bibfnamefont{T.}~\bibnamefont{Deutsch}},\ and\ \bibinfo {author}
  {\bibfnamefont{X.}~\bibnamefont{Blase}},\ }%
  \bibfield{journal}{%
  \bibinfo {journal} {Phys. Rev. B}\ }%
  \textbf{\bibinfo {volume} {86}},\ \bibinfo {pages} {155315} (\bibinfo {month}
  {Oct}\ \bibinfo {year} {2012})\BibitemShut{NoStop}%
\bibitem{Serrano98}%
  \BibitemOpen
  \bibfield{author}{%
  \bibinfo {author} {\bibfnamefont{L.}~\bibnamefont{Serrano-Andres}}\ and\
  \bibinfo {author} {\bibfnamefont{M.~P.}\ \bibnamefont{Fuelscher}},\ }%
  \bibfield{journal}{%
  \bibinfo {journal} {J. Am. Chem. Soc.}\ }%
  \textbf{\bibinfo {volume} {120}},\ \bibinfo {pages} {10912} (\bibinfo {year}
  {1998})\BibitemShut{NoStop}%
\bibitem{Peach08}%
  \BibitemOpen
  \bibfield{author}{%
  \bibinfo {author} {\bibfnamefont{M.~J.~G.}\ \bibnamefont{Peach}}, \bibinfo
  {author} {\bibfnamefont{P.}~\bibnamefont{Benfield}}, \bibinfo {author}
  {\bibfnamefont{T.}~\bibnamefont{Helgaker}},\ and\ \bibinfo {author}
  {\bibfnamefont{D.~J.}\ \bibnamefont{Tozer}},\ }%
  \bibfield{journal}{%
  \bibinfo {journal} {J. Chem. Phys.}\ }%
  \textbf{\bibinfo {volume} {128}},\ \bibinfo {pages} {044118} (\bibinfo {year}
  {2008})\BibitemShut{NoStop}%
\bibitem{Rocca10}%
  \BibitemOpen
  \bibfield{author}{%
  \bibinfo {author} {\bibfnamefont{D.}~\bibnamefont{Rocca}}, \bibinfo {author}
  {\bibfnamefont{D.}~\bibnamefont{Lu}},\ and\ \bibinfo {author}
  {\bibfnamefont{G.}~\bibnamefont{Galli}},\ }%
  \bibfield{journal}{%
  \bibinfo {journal} {J. Chem. Phys.}\ }%
  \textbf{\bibinfo {volume} {133}},\ \bibinfo {pages} {164109} (\bibinfo {year}
  {2010})\BibitemShut{NoStop}%
\bibitem{Farid}%
  \BibitemOpen
  \bibfield{author}{%
  \bibinfo {author} {\bibfnamefont{B.}~\bibnamefont{Farid}}, \bibinfo {author}
  {\bibfnamefont{R.}~\bibnamefont{Daling}}, \bibinfo {author}
  {\bibfnamefont{D.}~\bibnamefont{Lenstra}},\ and\ \bibinfo {author}
  {\bibfnamefont{W.}~\bibnamefont{van Haeringen}},\ }%
  \bibfield{journal}{%
  \bibinfo {journal} {Phys. Rev. B}\ }%
  \textbf{\bibinfo {volume} {38}},\ \bibinfo {pages} {7530} (\bibinfo {year}
  {1988})\BibitemShut{NoStop}%
\bibitem{Tiago05}%
  \BibitemOpen
  \bibfield{author}{%
  \bibinfo {author} {\bibfnamefont{M.~L.}\ \bibnamefont{Tiago}}\ and\ \bibinfo
  {author} {\bibfnamefont{J.~R.}\ \bibnamefont{Chelikowsky}},\ }%
  \bibfield{journal}{%
  \bibinfo {journal} {Solid State Communications}\ }%
  \textbf{\bibinfo {volume} {136}},\ \bibinfo {pages} {333 } (\bibinfo {year}
  {2005})\BibitemShut{NoStop}%
\bibitem{Dori06}%
  \BibitemOpen
  \bibfield{author}{%
  \bibinfo {author} {\bibfnamefont{N.}~\bibnamefont{Dori}}, \bibinfo {author}
  {\bibfnamefont{M.}~\bibnamefont{Menon}}, \bibinfo {author}
  {\bibfnamefont{L.}~\bibnamefont{Kilian}}, \bibinfo {author}
  {\bibfnamefont{M.}~\bibnamefont{Sokolowski}}, \bibinfo {author}
  {\bibfnamefont{L.}~\bibnamefont{Kronik}},\ and\ \bibinfo {author}
  {\bibfnamefont{E.}~\bibnamefont{Umbach}},\ }%
  \bibfield{journal}{%
  \bibinfo {journal} {Phys. Rev. B}\ }%
  \textbf{\bibinfo {volume} {73}},\ \bibinfo {pages} {195208} (\bibinfo {month}
  {May}\ \bibinfo {year} {2006})\BibitemShut{NoStop}%
\bibitem{Palummo09}%
  \BibitemOpen
  \bibfield{author}{%
  \bibinfo {author} {\bibfnamefont{M.}~\bibnamefont{Palummo}}, \bibinfo
  {author} {\bibfnamefont{C.}~\bibnamefont{Hogan}}, \bibinfo {author}
  {\bibfnamefont{F.}~\bibnamefont{Sottile}}, \bibinfo {author}
  {\bibfnamefont{P.}~\bibnamefont{Bagala}},\ and\ \bibinfo {author}
  {\bibfnamefont{A.}~\bibnamefont{Rubio}},\ }%
  \bibfield{journal}{%
  \bibinfo {journal} {J. Chem. Phys.}\ }%
  \textbf{\bibinfo {volume} {131}},\ \bibinfo {pages} {084102} (\bibinfo {year}
  {2009})\BibitemShut{NoStop}%
\bibitem{Marom11}%
  \BibitemOpen
  \bibfield{author}{%
  \bibinfo {author} {\bibfnamefont{N.}~\bibnamefont{Marom}}, \bibinfo {author}
  {\bibfnamefont{X.}~\bibnamefont{Ren}}, \bibinfo {author}
  {\bibfnamefont{J.~E.}\ \bibnamefont{Moussa}}, \bibinfo {author}
  {\bibfnamefont{J.~R.}\ \bibnamefont{Chelikowsky}},\ and\ \bibinfo {author}
  {\bibfnamefont{L.}~\bibnamefont{Kronik}},\ }%
  \bibfield{journal}{%
  \bibinfo {journal} {Phys. Rev. B}\ }%
  \textbf{\bibinfo {volume} {84}},\ \bibinfo {pages} {195143} (\bibinfo {year}
  {2011})\BibitemShut{NoStop}%
\bibitem{Foerster11}%
  \BibitemOpen
  \bibfield{author}{%
  \bibinfo {author} {\bibfnamefont{D.}~\bibnamefont{Foerster}}, \bibinfo
  {author} {\bibfnamefont{P.}~\bibnamefont{Koval}},\ and\ \bibinfo {author}
  {\bibfnamefont{D.}~\bibnamefont{Sanchez-Portal}},\ }%
  \bibfield{journal}{%
  \bibinfo {journal} {J. Chem. Phys.}\ }%
  \textbf{\bibinfo {volume} {135}},\ \bibinfo {pages} {074105} (\bibinfo {year}
  {2011})\BibitemShut{NoStop}%
\bibitem{Samsonidze11}%
  \BibitemOpen
  \bibfield{author}{%
  \bibinfo {author} {\bibfnamefont{G.}~\bibnamefont{Samsonidze}}, \bibinfo
  {author} {\bibfnamefont{M.}~\bibnamefont{Jain}}, \bibinfo {author}
  {\bibfnamefont{J.}~\bibnamefont{Deslippe}}, \bibinfo {author}
  {\bibfnamefont{M.~L.}\ \bibnamefont{Cohen}},\ and\ \bibinfo {author}
  {\bibfnamefont{S.~G.}\ \bibnamefont{Louie}},\ }%
  \bibfield{journal}{%
  \bibinfo {journal} {Phys. Rev. Lett.}\ }%
  \textbf{\bibinfo {volume} {107}},\ \bibinfo {pages} {186404} (\bibinfo {year}
  {2011})\BibitemShut{NoStop}%
\bibitem{Sharifzadeh12}%
  \BibitemOpen
  \bibfield{author}{%
  \bibinfo {author} {\bibfnamefont{S.}~\bibnamefont{Sharifzadeh}}, \bibinfo
  {author} {\bibfnamefont{I.}~\bibnamefont{Tamblyn}}, \bibinfo {author}
  {\bibfnamefont{P.}~\bibnamefont{Doak}}, \bibinfo {author}
  {\bibfnamefont{P.}~\bibnamefont{Darancet}},\ and\ \bibinfo {author}
  {\bibfnamefont{J.}~\bibnamefont{Neaton}},\ }%
  \bibfield{journal}{%
  \bibinfo {journal} {Euro. Phys. J. B}\ }%
  \textbf{\bibinfo {volume} {85}},\ \bibinfo {pages} {1} (\bibinfo {year}
  {2012})\BibitemShut{NoStop}%
\bibitem{Umari12}%
  \BibitemOpen
  \bibfield{author}{%
  \bibinfo {author} {\bibfnamefont{P.}~\bibnamefont{Umari}}\ and\ \bibinfo
  {author} {\bibfnamefont{S.}~\bibnamefont{Fabris}},\ }%
  \bibfield{journal}{%
  \bibinfo {journal} {J. Chem. Phys.}\ }%
  \textbf{\bibinfo {volume} {136}},\ \bibinfo {pages} {174310} (\bibinfo {year}
  {2012})\BibitemShut{NoStop}%
\bibitem{Marom12}%
  \BibitemOpen
  \bibfield{author}{%
  \bibinfo {author} {\bibfnamefont{N.}~\bibnamefont{Marom}}, \bibinfo {author}
  {\bibfnamefont{F.}~\bibnamefont{Caruso}}, \bibinfo {author}
  {\bibfnamefont{X.}~\bibnamefont{Ren}}, \bibinfo {author}
  {\bibfnamefont{O.~T.}\ \bibnamefont{Hofmann}}, \bibinfo {author}
  {\bibfnamefont{T.}~\bibnamefont{K\"orzd\"orfer}}, \bibinfo {author}
  {\bibfnamefont{J.~R.}\ \bibnamefont{Chelikowsky}}, \bibinfo {author}
  {\bibfnamefont{A.}~\bibnamefont{Rubio}}, \bibinfo {author}
  {\bibfnamefont{M.}~\bibnamefont{Scheffler}},\ and\ \bibinfo {author}
  {\bibfnamefont{P.}~\bibnamefont{Rinke}},\ }%
  \bibfield{journal}{%
  \bibinfo {journal} {Phys. Rev. B}\ }%
  \textbf{\bibinfo {volume} {86}},\ \bibinfo {pages} {245127} (\bibinfo {year}
  {2012})\BibitemShut{NoStop}%
\bibitem{Koerzdoerfer12}%
  \BibitemOpen
  \bibfield{author}{%
  \bibinfo {author} {\bibfnamefont{T.}~\bibnamefont{K\"orzd\"orfer}}\ and\
  \bibinfo {author} {\bibfnamefont{N.}~\bibnamefont{Marom}},\ }%
  \bibfield{journal}{%
  \bibinfo {journal} {Phys. Rev. B}\ }%
  \textbf{\bibinfo {volume} {86}},\ \bibinfo {pages} {041110} (\bibinfo {year}
  {2012})\BibitemShut{NoStop}%
\bibitem{Pham13}%
  \BibitemOpen
  \bibfield{author}{%
  \bibinfo {author} {\bibfnamefont{T.~A.}\ \bibnamefont{Pham}}, \bibinfo
  {author} {\bibfnamefont{H.-V.}\ \bibnamefont{Nguyen}}, \bibinfo {author}
  {\bibfnamefont{D.}~\bibnamefont{Rocca}},\ and\ \bibinfo {author}
  {\bibfnamefont{G.}~\bibnamefont{Galli}},\ }%
  \bibfield{journal}{%
  \bibinfo {journal} {Phys. Rev. B}\ }%
  \textbf{\bibinfo {volume} {87}},\ \bibinfo {pages} {155148} (\bibinfo {year}
  {2013})\BibitemShut{NoStop}%
\bibitem{Blase_FIESTA_code}%
  \BibitemOpen
  \bibfield{author}{%
  \bibinfo {author} {\bibfnamefont{X.}~\bibnamefont{Blase}}, \bibinfo {author}
  {\bibfnamefont{C.}~\bibnamefont{Attaccalite}},\ and\ \bibinfo {author}
  {\bibfnamefont{V.}~\bibnamefont{Olevano}},\ }%
  \bibfield{journal}{%
  \bibinfo {journal} {Phys. Rev. B}\ }%
  \textbf{\bibinfo {volume} {83}},\ \bibinfo {pages} {115103} (\bibinfo {year}
  {2011})\BibitemShut{NoStop}%
\bibitem{Faber_DNA}%
  \BibitemOpen
  \bibfield{author}{%
  \bibinfo {author} {\bibfnamefont{C.}~\bibnamefont{Faber}}, \bibinfo {author}
  {\bibfnamefont{C.}~\bibnamefont{Attaccalite}}, \bibinfo {author}
  {\bibfnamefont{V.}~\bibnamefont{Olevano}}, \bibinfo {author}
  {\bibfnamefont{E.}~\bibnamefont{Runge}},\ and\ \bibinfo {author}
  {\bibfnamefont{X.}~\bibnamefont{Blase}},\ }%
  \bibfield{journal}{%
  \bibinfo {journal} {Phys. Rev. B}\ }%
  \textbf{\bibinfo {volume} {83}},\ \bibinfo {pages} {115123} (\bibinfo {year}
  {2011})\BibitemShut{NoStop}%
\bibitem{FaridGW}%
  \BibitemOpen
  \bibfield{author}{%
  \bibinfo {author} {\bibfnamefont{A.}~\bibnamefont{Savin}},\ }%
  in\ \emph{\bibinfo {booktitle} {Electron Correlation in the Solid State}},\
  \bibinfo {editor} {edited by\ \bibinfo {editor}
  {\bibfnamefont{N.}~\bibnamefont{March}}}\ (\bibinfo {publisher} {World
  Scientific, Singapore},\ \bibinfo {year} {1999})\BibitemShut{NoStop}%
\bibitem{Vahtras1993514}%
  \BibitemOpen
  \bibfield{author}{%
  \bibinfo {author} {\bibfnamefont{O.}~\bibnamefont{Vahtras}}, \bibinfo
  {author} {\bibfnamefont{J.}~\bibnamefont{Alml$\ddot{\mathrm{o}}$f}},\ and\
  \bibinfo {author} {\bibfnamefont{M.}~\bibnamefont{Feyereisen}},\ }%
  \bibfield{journal}{%
  \bibinfo {journal} {Chemical Physics Letters}\ }%
  \textbf{\bibinfo {volume} {213}},\ \bibinfo {pages} {514 } (\bibinfo {year}
  {1993})\BibitemShut{NoStop}%
\bibitem{1367-2630-14-5-053020}%
  \BibitemOpen
  \bibfield{author}{%
  \bibinfo {author} {\bibfnamefont{X.}~\bibnamefont{Ren}}, \bibinfo {author}
  {\bibfnamefont{P.}~\bibnamefont{Rinke}}, \bibinfo {author}
  {\bibfnamefont{V.}~\bibnamefont{Blum}}, \bibinfo {author}
  {\bibfnamefont{J.}~\bibnamefont{Wieferink}}, \bibinfo {author}
  {\bibfnamefont{A.}~\bibnamefont{Tkatchenko}}, \bibinfo {author}
  {\bibfnamefont{A.}~\bibnamefont{Sanfilippo}}, \bibinfo {author}
  {\bibfnamefont{K.}~\bibnamefont{Reuter}},\ and\ \bibinfo {author}
  {\bibfnamefont{M.}~\bibnamefont{Scheffler}},\ }%
  \bibfield{journal}{%
  \bibinfo {journal} {New Journal of Physics}\ }%
  \textbf{\bibinfo {volume} {14}},\ \bibinfo {pages} {053020} (\bibinfo {year}
  {2012})\BibitemShut{NoStop}%
\bibitem{RI-MP2}%
  \BibitemOpen
  \bibfield{author}{%
  \bibinfo {author} {\bibfnamefont{F.}~\bibnamefont{Weigend}}\ and\ \bibinfo
  {author} {\bibfnamefont{M.}~\bibnamefont{Häser}},\ }%
  \bibfield{journal}{%
  \bibinfo {journal} {Theoretical Chemistry Accounts}\ }%
  \textbf{\bibinfo {volume} {97}},\ \bibinfo {pages} {331} (\bibinfo {year}
  {1997})\BibitemShut{NoStop}%
\bibitem{Cherkes09}%
  \BibitemOpen
  \bibfield{author}{%
  \bibinfo {author} {\bibfnamefont{I.}~\bibnamefont{Cherkes}}, \bibinfo
  {author} {\bibfnamefont{S.}~\bibnamefont{Klaiman}},\ and\ \bibinfo {author}
  {\bibfnamefont{N.}~\bibnamefont{Moiseyev}},\ }%
  \bibfield{journal}{%
  \bibinfo {journal} {Int. J. Quant. Chem.}\ }%
  \textbf{\bibinfo {volume} {109}},\ \bibinfo {pages} {2996} (\bibinfo {year}
  {2009})\BibitemShut{NoStop}%
\bibitem{Siesta}%
  \BibitemOpen
  \bibfield{author}{%
  \bibinfo {author} {\bibfnamefont{J.~M.}\ \bibnamefont{Soler}}, \bibinfo
  {author} {\bibfnamefont{E.}~\bibnamefont{Artacho}}, \bibinfo {author}
  {\bibfnamefont{J.~D.}\ \bibnamefont{Gale}}, \bibinfo {author}
  {\bibfnamefont{A.}~\bibnamefont{Garc\'ia}}, \bibinfo {author}
  {\bibfnamefont{J.}~\bibnamefont{Junquera}}, \bibinfo {author}
  {\bibfnamefont{P.}~\bibnamefont{Orde\'on}},\ and\ \bibinfo {author}
  {\bibfnamefont{D.}~\bibnamefont{Sanchez-Portal}},\ }%
  \bibfield{journal}{%
  \bibinfo {journal} {J. Phys.: Condens. Matter}\ }%
  \textbf{\bibinfo {volume} {14}},\ \bibinfo {pages} {2745} (\bibinfo {year}
  {2002})\BibitemShut{NoStop}%
\bibitem{LDA_1}%
  \BibitemOpen
  \bibfield{author}{%
  \bibinfo {author} {\bibfnamefont{D.}~\bibnamefont{Ceperley}}\ and\ \bibinfo
  {author} {\bibfnamefont{B.}~\bibnamefont{Alder}},\ }%
  \bibfield{journal}{%
  \bibinfo {journal} {Phys. Rev. Lett.}\ }%
  \textbf{\bibinfo {volume} {45}},\ \bibinfo {pages} {566} (\bibinfo {year}
  {1980})\BibitemShut{NoStop}%
\bibitem{LDA_3}%
  \BibitemOpen
  \bibfield{author}{%
  \bibinfo {author} {\bibfnamefont{S.~H.}\ \bibnamefont{Vosko}}, \bibinfo
  {author} {\bibfnamefont{L.}~\bibnamefont{Wilk}},\ and\ \bibinfo {author}
  {\bibfnamefont{M.}~\bibnamefont{Nusair}},\ }%
  \bibfield{journal}{%
  \bibinfo {journal} {Can. J. Phys.}\ }%
  \textbf{\bibinfo {volume} {58}},\ \bibinfo {pages} {1200} (\bibinfo {year}
  {1980})\BibitemShut{NoStop}%
\bibitem{LDA_2}%
  \BibitemOpen
  \bibfield{author}{%
  \bibinfo {author} {\bibfnamefont{J.~P.}\ \bibnamefont{Perdew}}\ and\ \bibinfo
  {author} {\bibfnamefont{A.}~\bibnamefont{Zunger}},\ }%
  \bibfield{journal}{%
  \bibinfo {journal} {Phys. Rev. B}\ }%
  \textbf{\bibinfo {volume} {23}},\ \bibinfo {pages} {5048} (\bibinfo {year}
  {1981})\BibitemShut{NoStop}%
\bibitem{Pseudos_Troullier_Martins}%
  \BibitemOpen
  \bibfield{author}{%
  \bibinfo {author} {\bibfnamefont{N.}~\bibnamefont{Troullier}}\ and\ \bibinfo
  {author} {\bibfnamefont{J.~L.}\ \bibnamefont{Martins}},\ }%
  \bibfield{journal}{%
  \bibinfo {journal} {Phys. Rev. B}\ }%
  \textbf{\bibinfo {volume} {43}},\ \bibinfo {pages} {1993} (\bibinfo {month}
  {Jan}\ \bibinfo {year} {1991})\BibitemShut{NoStop}%
\bibitem{TZ2P}%
  \BibitemOpen
  \bibinfo {note} {Similarly to strategies developed for post-Hartree-Fock
  correlated calculations, the first basis orbitals of the valence
  (\textit{s,p})-channels are taken to be the 2\textit{s} and 2\textit{p}
  eigenfunctions of isolated atoms in the corresponding pseudopotential
  approximation. The additional valence channels are taken to be two primitive
  Gaussians optimized to minimize the total energy at the DFT-LDA level. For
  carbon, the resulting most diffuse Gaussian present a decay coefficient
  $\alpha=$ 0.1 $\text{Bohr}^{-2}$, very close to the values optimized by
  Dunning at the cc-pVQZ level.\cite{Dunning89} Following
  Ref.~\onlinecite{Siesta}, the first \textit{d}-channel orbital is taken to be
  the polarization orbital of the atomic \textit{p} orbital, namely the
  \textit{d}-component of the perturbation induced by a uniform electric field,
  complemented by a primitive Gaussian with decay constant $\alpha=$ 0.3
  $\text{Bohr}^{-2}$ for carbon. Following Dunning, we finally add a single
  primitive Gaussian for the \textit{f}-channel (with e.g. $\alpha=$ 0.76
  $\text{Bohr}^{-2}$ for carbon).}\BibitemShut{Stop}%
\bibitem{Dunning89}%
  \BibitemOpen
  \bibfield{author}{%
  \bibinfo {author} {\bibfnamefont{T.~H.}\ \bibnamefont{Dunning}},\ }%
  \bibfield{journal}{%
  \bibinfo {journal} {J. Chem. Phys.}\ }%
  \textbf{\bibinfo {volume} {90}},\ \bibinfo {pages} {1007} (\bibinfo {year}
  {1989})\BibitemShut{NoStop}%
\bibitem{Bruneval13}%
  \BibitemOpen
  \bibfield{author}{%
  \bibinfo {author} {\bibfnamefont{F.}~\bibnamefont{Bruneval}}\ and\ \bibinfo
  {author} {\bibfnamefont{M.~A.~L.}\ \bibnamefont{Marques}},\ }%
  \bibfield{journal}{%
  \bibinfo {journal} {J. Chem. Theory Comput.}\ }%
  \textbf{\bibinfo {volume} {9}},\ \bibinfo {pages} {324} (\bibinfo {year}
  {2013})\BibitemShut{NoStop}%
\bibitem{Neaton12}%
  \BibitemOpen
  \bibfield{author}{%
  \bibinfo {author} {\bibfnamefont{S.}~\bibnamefont{Sharifzadeh}}, \bibinfo
  {author} {\bibfnamefont{A.}~\bibnamefont{Biller}}, \bibinfo {author}
  {\bibfnamefont{L.}~\bibnamefont{Kronik}},\ and\ \bibinfo {author}
  {\bibfnamefont{J.~B.}\ \bibnamefont{Neaton}},\ }%
  \bibfield{journal}{%
  \bibinfo {journal} {Phys. Rev. B}\ }%
  \textbf{\bibinfo {volume} {85}},\ \bibinfo {pages} {125307} (\bibinfo {year}
  {2012})\BibitemShut{NoStop}%
\bibitem{Evers12}%
  \BibitemOpen
  \bibfield{author}{%
  \bibinfo {author} {\bibfnamefont{M.~J.}\ \bibnamefont{van Setten}}, \bibinfo
  {author} {\bibfnamefont{F.}~\bibnamefont{Weigend}},\ and\ \bibinfo {author}
  {\bibfnamefont{F.}~\bibnamefont{Evers}},\ }%
  \bibfield{journal}{%
  \bibinfo {journal} {J. Chem. Theory Comput.}\ }%
  \textbf{\bibinfo {volume} {9}},\ \bibinfo {pages} {232} (\bibinfo {year}
  {2013})\BibitemShut{NoStop}%
\bibitem{Faber_C60}%
  \BibitemOpen
  \bibfield{author}{%
  \bibinfo {author} {\bibfnamefont{C.}~\bibnamefont{Faber}}, \bibinfo {author}
  {\bibfnamefont{J.~L.}\ \bibnamefont{Janssen}}, \bibinfo {author}
  {\bibfnamefont{M.}~\bibnamefont{C\^ot\'e}}, \bibinfo {author}
  {\bibfnamefont{E.}~\bibnamefont{Runge}},\ and\ \bibinfo {author}
  {\bibfnamefont{X.}~\bibnamefont{Blase}},\ }%
  \bibfield{journal}{%
  \bibinfo {journal} {Phys. Rev. B}\ }%
  \textbf{\bibinfo {volume} {84}},\ \bibinfo {pages} {155104} (\bibinfo {year}
  {2011})\BibitemShut{NoStop}%
\bibitem{Ciuchi12}%
  \BibitemOpen
  \bibfield{author}{%
  \bibinfo {author} {\bibfnamefont{S.}~\bibnamefont{Ciuchi}}, \bibinfo {author}
  {\bibfnamefont{R.~C.}\ \bibnamefont{Hatch}}, \bibinfo {author}
  {\bibfnamefont{H.}~\bibnamefont{H\"ochst}}, \bibinfo {author}
  {\bibfnamefont{C.}~\bibnamefont{Faber}}, \bibinfo {author}
  {\bibfnamefont{X.}~\bibnamefont{Blase}},\ and\ \bibinfo {author}
  {\bibfnamefont{S.}~\bibnamefont{Fratini}},\ }%
  \bibfield{journal}{%
  \bibinfo {journal} {Phys. Rev. Lett.}\ }%
  \textbf{\bibinfo {volume} {108}},\ \bibinfo {pages} {256401} (\bibinfo {year}
  {2012})\BibitemShut{NoStop}%
\bibitem{Bruneval06}%
  \BibitemOpen
  \bibfield{author}{%
  \bibinfo {author} {\bibfnamefont{F.}~\bibnamefont{Bruneval}}, \bibinfo
  {author} {\bibfnamefont{N.}~\bibnamefont{Vast}},\ and\ \bibinfo {author}
  {\bibfnamefont{L.}~\bibnamefont{Reining}},\ }%
  \bibfield{journal}{%
  \bibinfo {journal} {Phys. Rev. B}\ }%
  \textbf{\bibinfo {volume} {74}},\ \bibinfo {pages} {045102} (\bibinfo {year}
  {2006})\BibitemShut{NoStop}%
\bibitem{Mulliken}%
  \BibitemOpen
  \bibfield{author}{%
  \bibinfo {author} {\bibfnamefont{R.~S.}\ \bibnamefont{Mulliken}},\ }%
  \bibfield{journal}{%
  \bibinfo {journal} {Journal of the American Chemical Society}\ }%
  \textbf{\bibinfo {volume} {74}},\ \bibinfo {pages} {811} (\bibinfo {year}
  {1952})\BibitemShut{NoStop}%
\bibitem{Ma09}%
  \BibitemOpen
  \bibfield{author}{%
  \bibinfo {author} {\bibfnamefont{Y.}~\bibnamefont{Ma}}, \bibinfo {author}
  {\bibfnamefont{M.}~\bibnamefont{Rohlfing}},\ and\ \bibinfo {author}
  {\bibfnamefont{C.}~\bibnamefont{Molteni}},\ }%
  \bibfield{journal}{%
  \bibinfo {journal} {Phys. Rev. B}\ }%
  \textbf{\bibinfo {volume} {80}},\ \bibinfo {pages} {241405} (\bibinfo {year}
  {2009})\BibitemShut{NoStop}%
\bibitem{Gruening_TDA_09}%
  \BibitemOpen
  \bibfield{author}{%
  \bibinfo {author} {\bibfnamefont{M.}~\bibnamefont{Gr\"{u}ning}}, \bibinfo
  {author} {\bibfnamefont{A.}~\bibnamefont{Marini}},\ and\ \bibinfo {author}
  {\bibfnamefont{X.}~\bibnamefont{Gonze}},\ }%
  \bibfield{journal}{%
  \bibinfo {journal} {Nano Lett.}\ }%
  \textbf{\bibinfo {volume} {9}},\ \bibinfo {pages} {2820} (\bibinfo {year}
  {2009})\BibitemShut{NoStop}%
\bibitem{Blase_Duchemin_BSE_12}%
  \BibitemOpen
  \bibfield{author}{%
  \bibinfo {author} {\bibfnamefont{I.}~\bibnamefont{Duchemin}}, \bibinfo
  {author} {\bibfnamefont{T.}~\bibnamefont{Deutsch}},\ and\ \bibinfo {author}
  {\bibfnamefont{X.}~\bibnamefont{Blase}},\ }%
  \bibfield{journal}{%
  \bibinfo {journal} {Phys. Rev. Lett.}\ }%
  \textbf{\bibinfo {volume} {109}},\ \bibinfo {pages} {167801} (\bibinfo {year}
  {2012})\BibitemShut{NoStop}%
\bibitem{Note_NumberCondBands}%
  \BibitemOpen
  \bibinfo {note} {We carefully tested the influence of the number of involved
  conduction bands on the resulting excitation energies and oscillator
  strengths. Between 120 and 160 conduction bands, the excitation energies
  varied by less then 10 meV.}\BibitemShut{Stop}%
\bibitem{QuantumEspresso}%
  \BibitemOpen
  \bibfield{author}{%
  \bibinfo {author} {\bibfnamefont{P.}~\bibnamefont{Giannozzi}}\ and\ \bibinfo
  {author} {\bibnamefont{al.}},\ }%
  \bibfield{journal}{%
  \bibinfo {journal} {J. Phys. Condens. Matter}\ }%
  \textbf{\bibinfo {volume} {21}},\ \bibinfo {pages} {395502} (\bibinfo {year}
  {2009})\BibitemShut{NoStop}%
\bibitem{Yambo}%
  \BibitemOpen
  \bibfield{author}{%
  \bibinfo {author} {\bibfnamefont{A.}~\bibnamefont{Marini}}, \bibinfo {author}
  {\bibfnamefont{C.}~\bibnamefont{Hogan}}, \bibinfo {author}
  {\bibfnamefont{M.}~\bibnamefont{Gruning}},\ and\ \bibinfo {author}
  {\bibfnamefont{D.}~\bibnamefont{Varsano}},\ }%
  \bibfield{journal}{%
  \bibinfo {journal} {Comput. Phys. Comm.}\ }%
  \textbf{\bibinfo {volume} {180}},\ \bibinfo {pages} {1392} (\bibinfo {year}
  {2009})\BibitemShut{NoStop}%
\bibitem{Tozer99}%
  \BibitemOpen
  \bibfield{author}{%
  \bibinfo {author} {\bibfnamefont{D.~J.}\ \bibnamefont{Tozer}}, \bibinfo
  {author} {\bibfnamefont{R.~D.}\ \bibnamefont{Amos}}, \bibinfo {author}
  {\bibfnamefont{N.~C.}\ \bibnamefont{Handy}}, \bibinfo {author}
  {\bibfnamefont{B.~O.}\ \bibnamefont{Roos}},\ and\ \bibinfo {author}
  {\bibfnamefont{L.}~\bibnamefont{Serrano-Andres}},\ }%
  \bibfield{journal}{%
  \bibinfo {journal} {Mol. Phys.}\ }%
  \textbf{\bibinfo {volume} {97}},\ \bibinfo {pages} {859} (\bibinfo {year}
  {1999})\BibitemShut{NoStop}%
\bibitem{Akinaga09}%
  \BibitemOpen
  \bibfield{author}{%
  \bibinfo {author} {\bibfnamefont{Y.}~\bibnamefont{Akinaga}}\ and\ \bibinfo
  {author} {\bibfnamefont{S.}~\bibnamefont{Ten-No}},\ }%
  \bibfield{journal}{%
  \bibinfo {journal} {Int. J. Quant. Chem.}\ }%
  \textbf{\bibinfo {volume} {109}},\ \bibinfo {pages} {1905} (\bibinfo {year}
  {2009})\BibitemShut{NoStop}%
\bibitem{Structure_Fuelscher}%
  \BibitemOpen
  \bibinfo {note} {The structure we studied compares to structure 1a of
  Ref.~\onlinecite{Serrano98}, where results for different rotational
  structures of the dipeptide are presented.}\BibitemShut{Stop}%
\bibitem{B3lyp}%
  \BibitemOpen
  \bibfield{author}{%
  \bibinfo {author} {\bibfnamefont{A.~D.}\ \bibnamefont{Becke}},\ }%
  \bibfield{journal}{%
  \bibinfo {journal} {J. Chem. Phys.}\ }%
  \textbf{\bibinfo {volume} {98}},\ \bibinfo {pages} {1372} (\bibinfo {year}
  {1993})\BibitemShut{NoStop}%
\bibitem{Gaussian09}%
  \BibitemOpen
  \bibfield{author}{%
  \bibinfo {author} {\bibfnamefont{M.~J.}\ \bibnamefont{Frisch}}, \bibinfo
  {author} {\bibfnamefont{G.~W.}\ \bibnamefont{Trucks}}, \bibinfo {author}
  {\bibfnamefont{H.~B.}\ \bibnamefont{Schlegel}}, \bibinfo {author}
  {\bibfnamefont{G.~E.}\ \bibnamefont{Scuseria}}, \bibinfo {author}
  {\bibfnamefont{M.~A.}\ \bibnamefont{Robb}}, \bibinfo {author}
  {\bibfnamefont{J.~R.}\ \bibnamefont{Cheeseman}}, \bibinfo {author}
  {\bibfnamefont{G.}~\bibnamefont{Scalmani}}, \bibinfo {author}
  {\bibfnamefont{V.}~\bibnamefont{Barone}}, \bibinfo {author}
  {\bibfnamefont{B.}~\bibnamefont{Mennucci}}, \bibinfo {author}
  {\bibfnamefont{G.~A.}\ \bibnamefont{Petersson}}, \bibinfo {author}
  {\bibfnamefont{H.}~\bibnamefont{Nakatsuji}}, \bibinfo {author}
  {\bibfnamefont{M.}~\bibnamefont{Caricato}}, \bibinfo {author}
  {\bibfnamefont{X.}~\bibnamefont{Li}}, \bibinfo {author}
  {\bibfnamefont{H.~P.}\ \bibnamefont{Hratchian}}, \bibinfo {author}
  {\bibfnamefont{A.~F.}\ \bibnamefont{Izmaylov}}, \bibinfo {author}
  {\bibfnamefont{J.}~\bibnamefont{Bloino}}, \bibinfo {author}
  {\bibfnamefont{G.}~\bibnamefont{Zheng}}, \bibinfo {author}
  {\bibfnamefont{J.~L.}\ \bibnamefont{Sonnenberg}}, \bibinfo {author}
  {\bibfnamefont{M.}~\bibnamefont{Hada}}, \bibinfo {author}
  {\bibfnamefont{M.}~\bibnamefont{Ehara}}, \bibinfo {author}
  {\bibfnamefont{K.}~\bibnamefont{Toyota}}, \bibinfo {author}
  {\bibfnamefont{R.}~\bibnamefont{Fukuda}}, \bibinfo {author}
  {\bibfnamefont{J.}~\bibnamefont{Hasegawa}}, \bibinfo {author}
  {\bibfnamefont{M.}~\bibnamefont{Ishida}}, \bibinfo {author}
  {\bibfnamefont{T.}~\bibnamefont{Nakajima}}, \bibinfo {author}
  {\bibfnamefont{Y.}~\bibnamefont{Honda}}, \bibinfo {author}
  {\bibfnamefont{O.}~\bibnamefont{Kitao}}, \bibinfo {author}
  {\bibfnamefont{H.}~\bibnamefont{Nakai}}, \bibinfo {author}
  {\bibfnamefont{T.}~\bibnamefont{Vreven}}, \bibinfo {author}
  {\bibfnamefont{J.~A.}\ \bibnamefont{Montgomery}}, \bibinfo {author}
  {\bibfnamefont{J.~E.}\ \bibnamefont{Peralta}}, \bibinfo {author}
  {\bibfnamefont{F.}~\bibnamefont{Ogliaro}}, \bibinfo {author}
  {\bibfnamefont{M.}~\bibnamefont{Bearpark}}, \bibinfo {author}
  {\bibfnamefont{J.~J.}\ \bibnamefont{Heyd}}, \bibinfo {author}
  {\bibfnamefont{E.}~\bibnamefont{Brothers}}, \bibinfo {author}
  {\bibfnamefont{K.~N.}\ \bibnamefont{Kudin}}, \bibinfo {author}
  {\bibfnamefont{V.~N.}\ \bibnamefont{Staroverov}}, \bibinfo {author}
  {\bibfnamefont{R.}~\bibnamefont{Kobayashi}}, \bibinfo {author}
  {\bibfnamefont{J.}~\bibnamefont{Normand}}, \bibinfo {author}
  {\bibfnamefont{K.}~\bibnamefont{Raghavachari}}, \bibinfo {author}
  {\bibfnamefont{A.}~\bibnamefont{Rendell}}, \bibinfo {author}
  {\bibfnamefont{J.~C.}\ \bibnamefont{Burant}}, \bibinfo {author}
  {\bibfnamefont{S.~S.}\ \bibnamefont{Iyengar}}, \bibinfo {author}
  {\bibfnamefont{J.}~\bibnamefont{Tomasi}}, \bibinfo {author}
  {\bibfnamefont{M.}~\bibnamefont{Cossi}}, \bibinfo {author}
  {\bibfnamefont{N.}~\bibnamefont{Rega}}, \bibinfo {author}
  {\bibfnamefont{J.~M.}\ \bibnamefont{Millam}}, \bibinfo {author}
  {\bibfnamefont{M.}~\bibnamefont{Klene}}, \bibinfo {author}
  {\bibfnamefont{J.~E.}\ \bibnamefont{Knox}}, \bibinfo {author}
  {\bibfnamefont{J.~B.}\ \bibnamefont{Cross}}, \bibinfo {author}
  {\bibfnamefont{V.}~\bibnamefont{Bakken}}, \bibinfo {author}
  {\bibfnamefont{C.}~\bibnamefont{Adamo}}, \bibinfo {author}
  {\bibfnamefont{J.}~\bibnamefont{Jaramillo}}, \bibinfo {author}
  {\bibfnamefont{R.}~\bibnamefont{Gomperts}}, \bibinfo {author}
  {\bibfnamefont{R.~E.}\ \bibnamefont{Stratmann}}, \bibinfo {author}
  {\bibfnamefont{O.}~\bibnamefont{Yazyev}}, \bibinfo {author}
  {\bibfnamefont{A.~J.}\ \bibnamefont{Austin}}, \bibinfo {author}
  {\bibfnamefont{R.}~\bibnamefont{Cammi}}, \bibinfo {author}
  {\bibfnamefont{C.}~\bibnamefont{Pomelli}}, \bibinfo {author}
  {\bibfnamefont{J.~W.}\ \bibnamefont{Ochterski}}, \bibinfo {author}
  {\bibfnamefont{R.~L.}\ \bibnamefont{Martin}}, \bibinfo {author}
  {\bibfnamefont{K.}~\bibnamefont{Morokuma}}, \bibinfo {author}
  {\bibfnamefont{V.~G.}\ \bibnamefont{Zakrzewski}}, \bibinfo {author}
  {\bibfnamefont{G.~A.}\ \bibnamefont{Voth}}, \bibinfo {author}
  {\bibfnamefont{P.}~\bibnamefont{Salvador}}, \bibinfo {author}
  {\bibfnamefont{J.~J.}\ \bibnamefont{Dannenberg}}, \bibinfo {author}
  {\bibfnamefont{S.}~\bibnamefont{Dapprich}}, \bibinfo {author}
  {\bibfnamefont{A.~D.}\ \bibnamefont{Daniels}}, \bibinfo {author}
  {\bibfnamefont{Ö.}~\bibnamefont{Farkas}}, \bibinfo {author}
  {\bibfnamefont{J.~B.}\ \bibnamefont{Foresman}}, \bibinfo {author}
  {\bibfnamefont{J.~V.}\ \bibnamefont{Ortiz}}, \bibinfo {author}
  {\bibfnamefont{J.}~\bibnamefont{Cioslowski}},\ and\ \bibinfo {author}
  {\bibfnamefont{D.~J.}\ \bibnamefont{Fox}},\ }%
  \enquote{\bibinfo {title} {Gaussian~09 {R}evision {A}.1},}\ \bibinfo {note}
  {Gaussian Inc. Wallingford CT 2009}\BibitemShut{NoStop}%
\bibitem{Umari_DNA_11}%
  \BibitemOpen
  \bibfield{author}{%
  \bibinfo {author} {\bibfnamefont{X.}~\bibnamefont{Qian}}, \bibinfo {author}
  {\bibfnamefont{P.}~\bibnamefont{Umari}},\ and\ \bibinfo {author}
  {\bibfnamefont{N.}~\bibnamefont{Marzari}},\ }%
  \bibfield{journal}{%
  \bibinfo {journal} {Phys. Rev. B}\ }%
  \textbf{\bibinfo {volume} {84}},\ \bibinfo {pages} {075103} (\bibinfo {year}
  {2011})\BibitemShut{NoStop}%
\bibitem{Note_scissor}%
  \BibitemOpen
  \bibinfo {note} {The HOMO-LUMO gap used to start the BSE calculations in Ref.
  \onlinecite{Rocca10} was opened to enforce the agreement between the TDLDA
  and BSE excitation energies for the local $W1$ transition. Due to the
  iterative methodology used in this previous work, the $CTa$ transition with
  vanishing oscillator strength could not be obtained at the BSE level beyond
  the TDA.}\BibitemShut{Stop}%
\bibitem{Note_TDA2}%
  \BibitemOpen
  \bibinfo {note} {The use of the Tamm-Dancoff approximation at the $GW$/BSE
  level leads to increased excitation energies and a deteriorated spectrum as
  compared to CASPT2. In agreement with the 0.15 eV blue shift reported by
  Rocca and coworkers,\cite{Rocca10} the largest TDA induced shift concerns the
  $CTb$ excitation energy, which is blue-shifted by 0.17 eV in our
  calculations. The TDA further induces a small blue-shift of 0.03 eV for the
  $W1$ and $W2$ transitions, in perfect agreement with
  Ref.~\onlinecite{Rocca10}. The $CTa$ charge-transfer state is marginally
  affected by a 0.01 eV blue-shift.}\BibitemShut{Stop}%
\bibitem{Faleev04}%
  \BibitemOpen
  \bibfield{author}{%
  \bibinfo {author} {\bibfnamefont{S.~V.}\ \bibnamefont{Faleev}}, \bibinfo
  {author} {\bibfnamefont{M.}~\bibnamefont{van Schilfgaarde}},\ and\ \bibinfo
  {author} {\bibfnamefont{T.}~\bibnamefont{Kotani}},\ }%
  \bibfield{journal}{%
  \bibinfo {journal} {Phys. Rev. Lett.}\ }%
  \textbf{\bibinfo {volume} {93}},\ \bibinfo {pages} {126406} (\bibinfo {year}
  {2004})\BibitemShut{NoStop}%
\bibitem{Shishkin07}%
  \BibitemOpen
  \bibfield{author}{%
  \bibinfo {author} {\bibfnamefont{M.}~\bibnamefont{Shishkin}}\ and\ \bibinfo
  {author} {\bibfnamefont{G.}~\bibnamefont{Kresse}},\ }%
  \bibfield{journal}{%
  \bibinfo {journal} {Phys. Rev. B}\ }%
  \textbf{\bibinfo {volume} {75}},\ \bibinfo {pages} {235102} (\bibinfo {year}
  {2007})\BibitemShut{NoStop}%
\bibitem{Rostgaard10}%
  \BibitemOpen
  \bibfield{author}{%
  \bibinfo {author} {\bibfnamefont{C.}~\bibnamefont{Rostgaard}}, \bibinfo
  {author} {\bibfnamefont{K.~W.}\ \bibnamefont{Jacobsen}},\ and\ \bibinfo
  {author} {\bibfnamefont{K.~S.}\ \bibnamefont{Thygesen}},\ }%
  \bibfield{journal}{%
  \bibinfo {journal} {Phys. Rev. B}\ }%
  \textbf{\bibinfo {volume} {81}},\ \bibinfo {pages} {085103} (\bibinfo {year}
  {2010})\BibitemShut{NoStop}%
\bibitem{Ke11}%
  \BibitemOpen
  \bibfield{author}{%
  \bibinfo {author} {\bibfnamefont{S.-H.}\ \bibnamefont{Ke}},\ }%
  \bibfield{journal}{%
  \bibinfo {journal} {Phys. Rev. B}\ }%
  \textbf{\bibinfo {volume} {84}},\ \bibinfo {pages} {205415} (\bibinfo {year}
  {2011})\BibitemShut{NoStop}%
\bibitem{Bruneval12}%
  \BibitemOpen
  \bibfield{author}{%
  \bibinfo {author} {\bibfnamefont{F.}~\bibnamefont{Bruneval}},\ }%
  \bibfield{journal}{%
  \bibinfo {journal} {J. Chem. Phys.}\ }%
  \textbf{\bibinfo {volume} {136}},\ \bibinfo {pages} {194107} (\bibinfo {year}
  {2012})\BibitemShut{NoStop}%
\bibitem{Gatti07}%
  \BibitemOpen
  \bibfield{author}{%
  \bibinfo {author} {\bibfnamefont{M.}~\bibnamefont{Gatti}}, \bibinfo {author}
  {\bibfnamefont{F.}~\bibnamefont{Bruneval}}, \bibinfo {author}
  {\bibfnamefont{V.}~\bibnamefont{Olevano}},\ and\ \bibinfo {author}
  {\bibfnamefont{L.}~\bibnamefont{Reining}},\ }%
  \bibfield{journal}{%
  \bibinfo {journal} {Phys. Rev. Lett.}\ }%
  \textbf{\bibinfo {volume} {99}},\ \bibinfo {pages} {266402} (\bibinfo {year}
  {2007})\BibitemShut{NoStop}%
\bibitem{Vidal10a}%
  \BibitemOpen
  \bibfield{author}{%
  \bibinfo {author} {\bibfnamefont{J.}~\bibnamefont{Vidal}}, \bibinfo {author}
  {\bibfnamefont{F.}~\bibnamefont{Trani}}, \bibinfo {author}
  {\bibfnamefont{F.}~\bibnamefont{Bruneval}}, \bibinfo {author}
  {\bibfnamefont{M.~A.~L.}\ \bibnamefont{Marques}},\ and\ \bibinfo {author}
  {\bibfnamefont{S.}~\bibnamefont{Botti}},\ }%
  \bibfield{journal}{%
  \bibinfo {journal} {Phys. Rev. Lett.}\ }%
  \textbf{\bibinfo {volume} {104}},\ \bibinfo {pages} {136401} (\bibinfo {year}
  {2010})\BibitemShut{NoStop}%
\bibitem{Vidal10b}%
  \BibitemOpen
  \bibfield{author}{%
  \bibinfo {author} {\bibfnamefont{J.}~\bibnamefont{Vidal}}, \bibinfo {author}
  {\bibfnamefont{S.}~\bibnamefont{Botti}}, \bibinfo {author}
  {\bibfnamefont{P.}~\bibnamefont{Olsson}}, \bibinfo {author}
  {\bibfnamefont{J.-F.}\ \bibnamefont{Guillemoles}},\ and\ \bibinfo {author}
  {\bibfnamefont{L.}~\bibnamefont{Reining}},\ }%
  \bibfield{journal}{%
  \bibinfo {journal} {Phys. Rev. Lett.}\ }%
  \textbf{\bibinfo {volume} {104}},\ \bibinfo {pages} {056401} (\bibinfo {year}
  {2010})\BibitemShut{NoStop}%
\bibitem{Rangel12}%
  \BibitemOpen
  \bibfield{author}{%
  \bibinfo {author} {\bibfnamefont{T.}~\bibnamefont{Rangel}}, \bibinfo {author}
  {\bibfnamefont{D.}~\bibnamefont{Kecik}}, \bibinfo {author}
  {\bibfnamefont{P.~E.}\ \bibnamefont{Trevisanutto}}, \bibinfo {author}
  {\bibfnamefont{G.-M.}\ \bibnamefont{Rignanese}}, \bibinfo {author}
  {\bibfnamefont{H.}~\bibnamefont{Van~Swygenhoven}},\ and\ \bibinfo {author}
  {\bibfnamefont{V.}~\bibnamefont{Olevano}},\ }%
  \bibfield{journal}{%
  \bibinfo {journal} {Phys. Rev. B}\ }%
  \textbf{\bibinfo {volume} {86}},\ \bibinfo {pages} {125125} (\bibinfo {year}
  {2012})\BibitemShut{NoStop}%
\bibitem{LUMOvalue}%
  \BibitemOpen
  \bibinfo {note} {The COHSEX, $GW@LDA$ and $GW@COHSEX$ LUMO energies are found
  to be 2.32 eV, 2.42 eV and 2.47 eV, respectively. This can be compared to the
  DFT-LDA starting Kohn-Sham value (-1.17 eV) and to the all-electron DFT-B3LYP
  6-311G(d,p) $\Delta$SCF value (2.13 eV) obtained with the Gaussian09
  code.}\BibitemShut{Stop}%
\bibitem{Sangalli11}%
  \BibitemOpen
  \bibfield{author}{%
  \bibinfo {author} {\bibfnamefont{D.}~\bibnamefont{Sangalli}}, \bibinfo
  {author} {\bibfnamefont{P.}~\bibnamefont{Romaniello}}, \bibinfo {author}
  {\bibfnamefont{G.}~\bibnamefont{Onida}},\ and\ \bibinfo {author}
  {\bibfnamefont{A.}~\bibnamefont{Marini}},\ }%
  \bibfield{journal}{%
  \bibinfo {journal} {J. Chem. Phys.}\ }%
  \textbf{\bibinfo {volume} {134}},\ \bibinfo {pages} {034115} (\bibinfo {year}
  {2011})\BibitemShut{NoStop}%
\end{thebibliography}

\end{document}